\documentclass[nofootinbib,amsmath,amssymb,aps,prd,10pt,floatfix,showkeys]{revtex4-2}%

\usepackage{dcolumn}%
\usepackage{bm}%
\usepackage{color}
\usepackage{lipsum}
\usepackage{multirow}
\usepackage{url}
\usepackage{hyperref}
\usepackage{makecell}
\usepackage{float}
\usepackage{soul}

\usepackage{mathtools}
\DeclareMathOperator{\sech}{sech}

\begin{document}

\title{Astrophysical positronium and Dicke superradiance}

\author{Abdaljalel E. Alizzi}
\email{abdaljalel90@gmail.com}
\affiliation{Budker Institute of Nuclear Physics, Novosibirsk 630 090, Russia}
\affiliation{Department of Physics, Novosibirsk State University, Novosibirsk 630 090, Russia}
\affiliation{Department of Physics, Al Furat University, Deir-ez-Zor, Syrian Arab Republic}

\author{Zurab~K.~Silagadze}
\email{Z.K.Silagadze@inp.nsk.su}
\affiliation{Budker Institute of Nuclear Physics, Novosibirsk 630 090, Russia}
\affiliation{Department of Physics, Novosibirsk State University, Novosibirsk 630 090, Russia}

\begin{abstract}
Dicke superradiance is a fascinating phenomenon in which a large number of atoms cooperate to produce a brief and very intense burst of spontaneous emission. This phenomenon has been well studied in the laboratory, but its astrophysical aspects have only recently attracted the attention of a small number of researchers. Since the phenomenon of Dicke superradiance is relatively little known to the wider astrophysical community, we provide a fairly detailed review of its elementary theory in the appendix and speculate on the significance of superradiance for astrophysical hydrogen and positronium, given the abundant formation of the latter near the galactic center.
\end{abstract}


\maketitle

\section{Introduction}
A year after Anderson's experimental discovery of the positron, Mohorovi\v{c}i\'{c}  \cite{Mohorovicic_1934} suggested the possibility of the existence of positronium (which he called
electrum) and proposed searching for astrophysical sources of positronium by detecting its spectral lines. Unfortunately, this suggestion was not appreciated or widely known, and was therefore not mentioned in several later papers that independently predicted the existence of what we now call positronium \cite{Cassidy_2018,Randic_2009,Kragh_1990}.

In astrophysical context, positronium in excited states cascades down to lower energy levels, emitting photons (recombination lines) at specific wavelengths. Its recombination lines are analogous to hydrogen's spectral lines but with frequencies roughly half those of hydrogen due to the reduced mass. These lines, particularly the optical Lyman-alpha line (2431 \AA) and radio recombination lines, offer a unique way to study positronium formation and annihilation processes in astrophysical environments.

Despite periodic renewed interest, no astrophysical recombination lines of positronium has been detected in the last 90 years since this hypothesis was put forward, probably due to the high background of the near-infrared Balmer lines and the strong attenuation of the ultraviolet Lyman lines due to Rayleigh scattering by dust in the interstellar medium \cite{Ellis_2017}. Observation of such recombination lines is particularly important given the half-century-long mystery of the origin of galactic 511 keV gamma rays measured by various instruments since their first observation in pioneering balloon experiments \cite{Siegert_2023,Prantzos_2011}, as it allows an angular resolution increase of $10^4$ compared to gamma-ray observations \cite{Ellis_2017}.

Positrons in space can be produced by a variety of mechanisms \cite{Prantzos_2011}:
\begin{itemize}
\item $\beta^+$-decay of unstable nuclei produced in stellar explosions, such as $^{56}Ni$, $^{22}Na$, $^{44}Ti$, and $^{26}Al$;
\item inelastic collisions of relativistic cosmic ray protons with interstellar gas and the production of secondary positrons in decays of positively charged mesons;
\item pair production in $\gamma-\gamma$ interactions;
\item rapidly rotating magnetized neutron stars (millisecond pulsars and magnetars), which can accelerate charged particles to Lorentz factors high enough to initiate cascade production of $e^+e^-$ pairs;
\item X-ray binaries and microquasars (for example, the microquasar 1E1740.7-2942, known as ``the Great Annihilator" \cite{Mirabel_1992});
\item pair production around the Galactic supermassive black hole.
\end{itemize}

The observed gamma-ray flux shows that in the Milky Way about $5\cdot 10^{43}$ positrons are annihilated every second near the galactic center. Moreover, the morphology of the 511 keV signal, emanating from the galactic bulge, center and disk with a bulge-to-disk luminosity ratio of $\sim 1$, is unique and does not resemble any other distribution of astrophysical sources (see Fig. 1 in \cite{Siegert_2023}).

A certain fraction of the astrophysical positrons produced, $f_{Ps}$, forms positronium either through radiative recombination or in an endoenergetic charge exchange reaction in which a positron captures an electron from an atom or molecule, forming positronium in flight if the kinetic energy of the positron is greater than the threshold energy of the charge exchange reaction \cite{Prantzos_2011,Ellis_2017}. This fraction can be estimated from the observed intensities of the two photon 511 keV line and the continuous emission of three photons with energies below 511 keV as follows \cite{Prantzos_2011,Leventhal_1992}. Positronium will be formed one-quarter of the time in the spin-singlet state of para-positronium and three-quarters of the time in the spin-triplet state of ortho-positronium. Charge conjugation symmetry forbids annihilation of $C$-even para-positronium into an odd number of photons and $C$-odd ortho-positronium  annihilating into an even number of photons. Therefore, ortho-positronium decays almost exclusively into three photons and thus produces a continuous spectrum of gamma rays with intensity $I_{cont}\sim 3\frac{3}{4}f_{Ps}=\frac{9}{4}f_{Ps}$, where the first factor of 3 arises because each decay produces three gamma rays. Para-positronium decays almost exclusively into two photons. However, another process of two photons production involves the direct annihilation of positrons with electrons when some $1-f_{Ps}$ positrons annihilate, without producing positronium. Therefore, the intensity of the 511 keV line is $I_{511}\sim 2\left(1-f_{Ps}+\frac{1}{4}f_{Ps}\right)=2\left(1-\frac{3}{4}f_{Ps}\right)$. Thus, we have $\frac{I_{511}}{I_{cont}}=\frac{8-6f_{Ps}}{9f_{Ps}}$, which implies
\begin{equation}
   f_{Ps}=\frac{8}{6+9\,\frac{I_{511}}{I_{cont}}}.
\label{eq1}   
\end{equation}
In fact, the observed intensities indicate that almost all positrons annihilate through the formation of positronium \cite{Siegert_2023}.

As we see, there is a region in the center of the Galaxy where positronium is present in large quantities. An interesting question is whether under such circumstances the cooperative coherence effects first considered by Dicke in his seminal paper \cite{Dicke_1954} can occur.  Our goal in this note is to speculate on this possibility. Readers unfamiliar with Dicke superradiance are advised to first consult the appendix, which provides a fairly detailed description of its elementary theory, as well as the basics of the hydrogen and positronium spin flip-lines.

\section{Dicke superradiance in astrophysics} 
The intensive research carried out in the quantum optics community following the first experimental confirmation of superradiance \cite{Skribanowitz_1973}, almost 20 years after the publication of Dicke's paper \cite{Dicke_1954}, has until recently been largely ignored in the astrophysical community, despite the discovery of astronomical masers in the interstellar medium \cite{Weaver_1965} indicating the possibility of such phenomena in an astrophysical context. The situation began to change after the possibility of superradiance in the spin-flip line of atomic hydrogen at 21 cm was discussed, and the application of superradiance to the 1612 MHz OH, 6.7 GHz CH3OH, and 22 GHz H2O maser lines was used to explain the observations of intensity bursts found in these spectral lines for some astronomical sources that cannot be easily explained within the maser theory \cite{Rajabi_2016-I,Rajabi_2016-II,Rajabi_2017,Houde_2018,Rajabi_2019,Houde_2019,Rajabi_2020,Houde_2022,Rajabi_2023,Rashidi_2024}. For example, periodic bursts in the methanol lines at 6.7 GHz and 12.2 GHz, as well as in the OH molecular transitions at 1665 MHz and 1667 MHz, observed in the star-forming region G9.62+0.20E, are best explained by Dicke superradiance, as shown by detailed modeling \cite{Rajabi_2023}.
The phenomenon of subradiance accompanying superradiance has even been considered in the context of dark matter \cite{Houde_2024}. In this work we focus on the spin-flip lines of atomic hydrogen and positronium.

Causality can limit the maximum number of atoms that can participate in a superradiance pulse. This maximum number can be estimated in the following way \cite{Eberly_1972,Arecchi_1970}. If a cylindrical sample of length $L$ and cross-section $S$ emits a superradiant pulse of characteristic time of the order of $T_{R}$, then for cooperative behavior to be possible, causality requires that the emitting atoms be within $cT_{R}$ distance of each other. Therefore, $L\le cT_{R}=2cT_{SR}$, where $T_{SR}$ is the duration of the  superradiant peak (For large samples the characteristic timescale of superradiance is still $T_{R}$ but the duration of the peak scales as $2\sqrt{T_{R} T_{D}}$ when dephasing is negligible \cite{Feld_1976}), and using $T_{SR}=\frac{4\pi S}{3N\lambda^2_\omega}\,\tau_0,$ where $\lambda_\omega=\frac{2\pi c}{\omega}$ (see (\ref{eq102}) in the appendix for how this was obtained) and assuming that the Fresnel number (\ref{eq90}) is unity, so that $S=2\pi Lc/\omega$, this Arecchi-Courtens criterion transforms into (where $\tau_0$ is the characteristic time of spontaneous decay):
\begin{equation}
 N\le \frac{4}{3}\omega\tau_0=\left \{ \begin{array}{l} 4.2\cdot 10^{24},\;\;\mathrm{for\; hydrogen},\\5.1\cdot 10^{19},\;\;\mathrm{for \;positronium}.\end{array}\right .   \label{eq138}
\end{equation}
It should be noted, however, that this limitation applies only when all atoms are inverted simultaneously (e.g., when the radiation responsible for the atomic population inversion propagates transversely to the inverted region). In this case, atoms located at a distance greater than $cT_R$ will participate in spatially distinct superradiance groups before they can interact with each other. However, this limitation does not apply when the inversion wave propagates longitudinally through the sample. In this case, there is no limit on the distance between atoms, since the excitation follows the superradiance signal along the sample (see \cite{Gross_1982} for a more detailed discussion).

According to (\ref{eq109}), the initial Bloch angles corresponding to the maximum numbers in (\ref{eq138}) are
\begin{equation}
 \theta_0=\left \{ \begin{array}{l} 9.8\cdot 10^{-13},\;\;\mathrm{for\; hydrogen},\\2.8\cdot10^{-10},\;\;\mathrm{for \;positronium}.\end{array}\right .   \label{eq139}
\end{equation}
We can use these values of $\theta_0$ in the initial conditions (\ref{eq109}) and solve the Burnham-Chiao equation (\ref{eq108}) numerically. For this, we used the DRKNYS \cite{CERNLIB} routine from the CERNLIB library, which solves second-order differential equations using the Runge–Kutta–Nystr\"{o}m method and outputs both the function $\theta(q)$ and its derivative $\frac{d\theta(q)}{dq}$.

\begin{figure}[ht]
    \centering
    \includegraphics[width=0.45\textwidth]{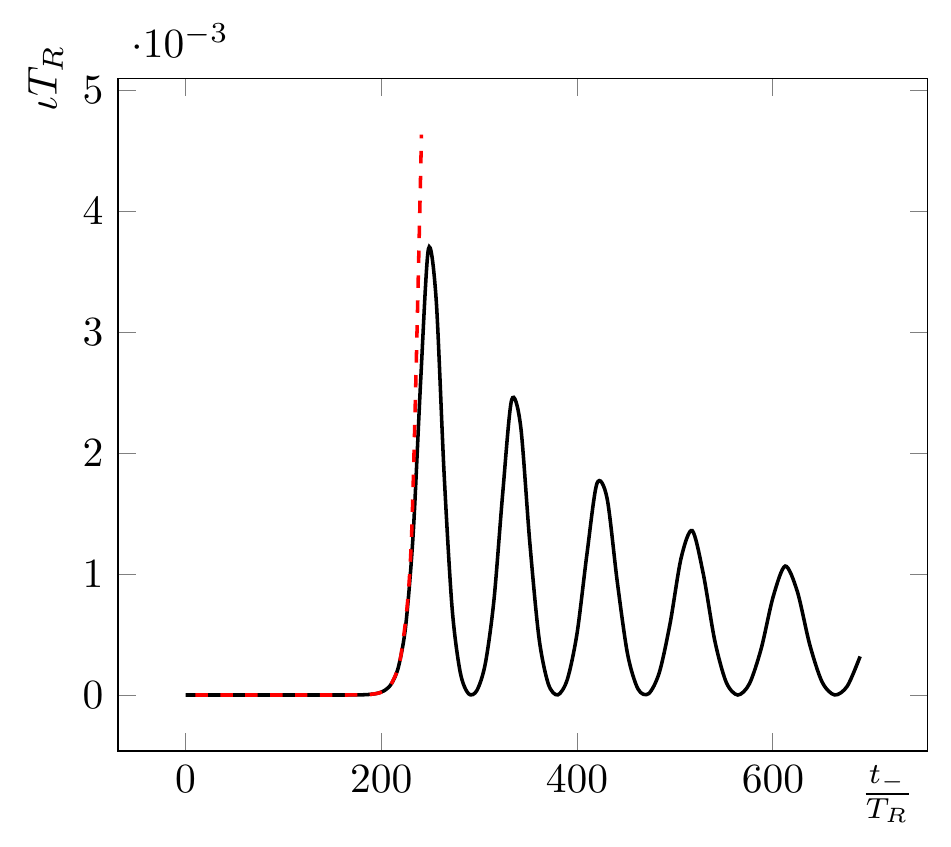}
    \includegraphics[width=0.45\textwidth]{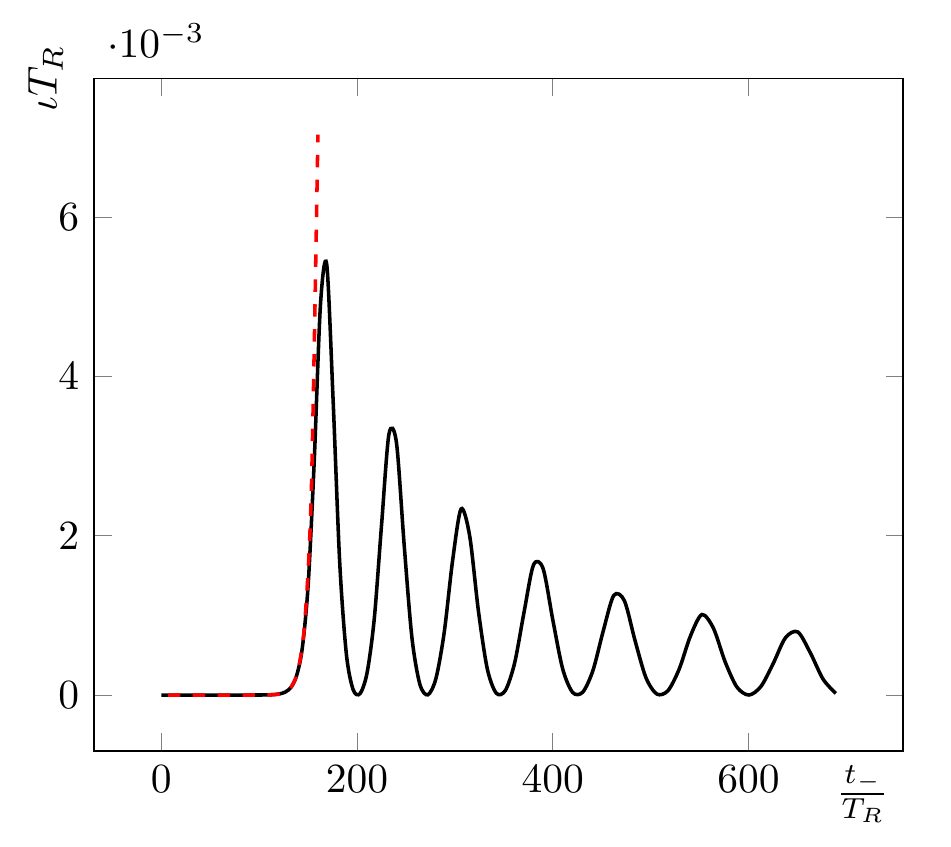}
     \caption{Radiance $\iota T_R$ of ideal samples of atoms of hydrogen (left) and of positronium (right) as a function of dimensionless time $t_-/T_R$. The number of atoms is maximal and corresponds to (\ref{eq138}). Red dashed line -- small Bloch angle approximation $\iota T_R\approx \theta_0^2 \,I_1(q)^2/q^2$.}
\label{Fig1}
\end{figure}

The intensity of superradiance can be conveniently characterized as the number of photons emitted from the end of the sample at $z=L$ per unit time and per atom \cite{Ermolaev_1996}, and this quantity $\iota$ can be expressed through the Poynting vector flux averaged over rapid oscillations of the electromagnetic field (hence the additional factor $1/2$ in the Poynting vector):
\begin{equation}
\iota=\frac{c}{8\pi}|E|^2\frac{S}{\hbar\omega N}=\frac{c}{4L}|\epsilon|^2,
\label{eq140}
\end{equation}
where in the last step we used (\ref{eq91}) to express $|E|^2$ in terms of $|\epsilon|^2$. But, according to (\ref{eq106}),
\begin{equation}
\epsilon=\frac{\partial \theta}{\partial\tau_-}=\sqrt{\frac{\xi}{\tau_-}}\,\frac{d\theta}{dq}=\sqrt{\frac{z}{c\,t_-}}\,\frac{d\theta}{dq},\;\;\;t_-=t-\frac{z}{c},
\label{eq141}
\end{equation}
and since $q=2\sqrt{\frac{t_-}{T_R}}$ when $z=L$, we get 
\begin{equation}
\iota T_R=\frac{1}{q^2}\left(\frac{d\theta}{dq}\right)^2.
\label{eq142}
\end{equation}
Fig. \ref{Fig1} shows the dimensionless radiance $\iota T_R$ versus dimensionless retarded time $t_-/T_R=q^2/4$. From the figure it is clear that the superradiance begins with some delay. This delay is natural since the system needs time to create a macroscopic dipole moment from a small initial quantum polarization fluctuation, and a small initial value of $\theta_0$ leads to a relatively large delay \cite{Ermolaev_1996}. The magnitude of the delay is not accurately described by the expression (\ref{eq101}), since this result was obtained for the limiting case of a small sample. A more accurate approximation can be obtained as follows \cite{Ermolaev_1996}.

When the Bloch angle $\theta$ is small, $\sin{\theta}\approx \theta$ and the solution of (\ref{eq108}) with the initial conditions (\ref{eq109}) in this approximation is $\theta(q)=\theta_0I_0(q)$, where $I_0$ is the modified Bessel function of the first kind and of zero order. Fig. \ref{Fig1} shows that the first peak of radiation occurs at a value of $q$, when the small-angle approximation begins to break down. Therefore, if the delay time $T_D$ is defined as the interval from the moment of population inversion of the sample to the moment of the appearance of the first peak of the emitted radiation \cite{Feld_1976}, then the corresponding $q$ is determined from the condition
\begin{equation}
\theta_0I_0(q)\approx 1.
\label{eq143}
\end{equation}
Since we expect $q\gg 1$ (long delay time), we can use the asymptotic expression
$I_0(q)\approx e^q/\sqrt{2\pi q}$ and rewrite (\ref{eq143}) as follows
\begin{equation}
    q=\ln{\frac{\sqrt{2\pi q}}{\theta_0}}=\ln{\frac{\sqrt{2\pi}}{\theta_0}}+\ln{\sqrt{q}}\,.
 \label{eq144}   
\end{equation}
Assuming that the first term on the right-hand side of (\ref{eq144}) is the leading term, we can solve (\ref{eq144}) approximately by writing $q=q^{(0)}+q^{(1)}$, $q^{(0)}=\ln{\frac{\sqrt{2\pi}}{\theta_0}}$, $q^{(1)}=\ln{\sqrt{q^{(0)}}}$, and obtain
\begin{equation}
q_D\approx\ln{\left(\frac{\sqrt{2\pi}}{\theta_0}\sqrt{\ln{\frac{\sqrt{2\pi}}{\theta_0}}}\right) },
 \label{eq145}
\end{equation}
and since $T_D$ can be identified with the retarded time corresponding to $z=L$ and $q=q_D$ (assuming $T_D\gg L/c$ and neglecting the time required for the population inversion), the final formula for $T_D$ takes the form
\begin{equation}
 T_D\approx\frac{T_R}{4}\,\ln^2{\left(\frac{\sqrt{2\pi}}{\theta_0}\sqrt{\ln{\frac{\sqrt{2\pi}}{\theta_0}}}\right) }.
 \label{eq146}
\end{equation}
For $\theta_0$-s from (\ref{eq139}), (\ref{eq145}) gives $T_D\approx 228.7\, T_R$ and $T_D\approx 149.8\,T_R$ for hydrogen and positronium, respectively. These numbers approximate $T_D$, which follows from the numerical solution of (\ref{eq143}), with an accuracy better than $0.3\%$.

Note that a slightly different estimate of $T_D$ by MacGillivray and Feld \cite{Feld_1976,Ermolaev_1996} is widely used in the literature:
\begin{equation}
 T_D\approx \frac{T_R}{4}\,\ln^2{\frac{\theta_0}{2\pi}} .
 \label{eq147}
\end{equation}
For ideal samples of hydrogen and positronium (\ref{eq147}) gives 
$T_D\approx\, 217.4\, T_R$ and $T_D\approx 142.0\,T_R$, respectively.

According to Fig. \ref{Fig1}, the energy stored in the inverted system is emitted in several successive bursts of decreasing amplitude, a phenomenon known as ringing. Ringing occurs when some of the energy released in a superradiant burst is absorbed by deactivated atoms, which are then re-excited and ready to initiate the next superradiant burst.

In fact, superradiance can be reduced or even stopped by any process that disrupts phase synchronization between atoms or incoherently reduces the population of their excited levels. These non-ideality effects can be taken into account phenomenologically by adding appropriate terms to the Maxwell-Bloch equations (\ref{eq89})
\cite{Rajabi_2016-I,Rajabi_2016-II}:
\begin{eqnarray} &&
\frac{\partial R}{\partial t}=-i\frac{d}{\hbar}\,EZ-\frac{R}{T_2},\;\;
\frac{\partial Z}{\partial t}=i\frac{d}{2\hbar}\left (ER^*-E^*R\right)-\frac{Z}{T_1},\nonumber \\ &&\frac{\partial E}{\partial z}+\frac{1}{c}\frac{\partial E}{\partial t}=i\frac{2\pi\omega \,n}{c}d\,R,
\label{eq148}
\end{eqnarray}
where $T_1$ is the characteristic time scale of the incoherent decay of the inverted population difference, and $T_2$ is the characteristic time scale of the violation of phase synchronization.

\begin{figure}[ht]
    \centering
    \includegraphics[width=0.7\textwidth]{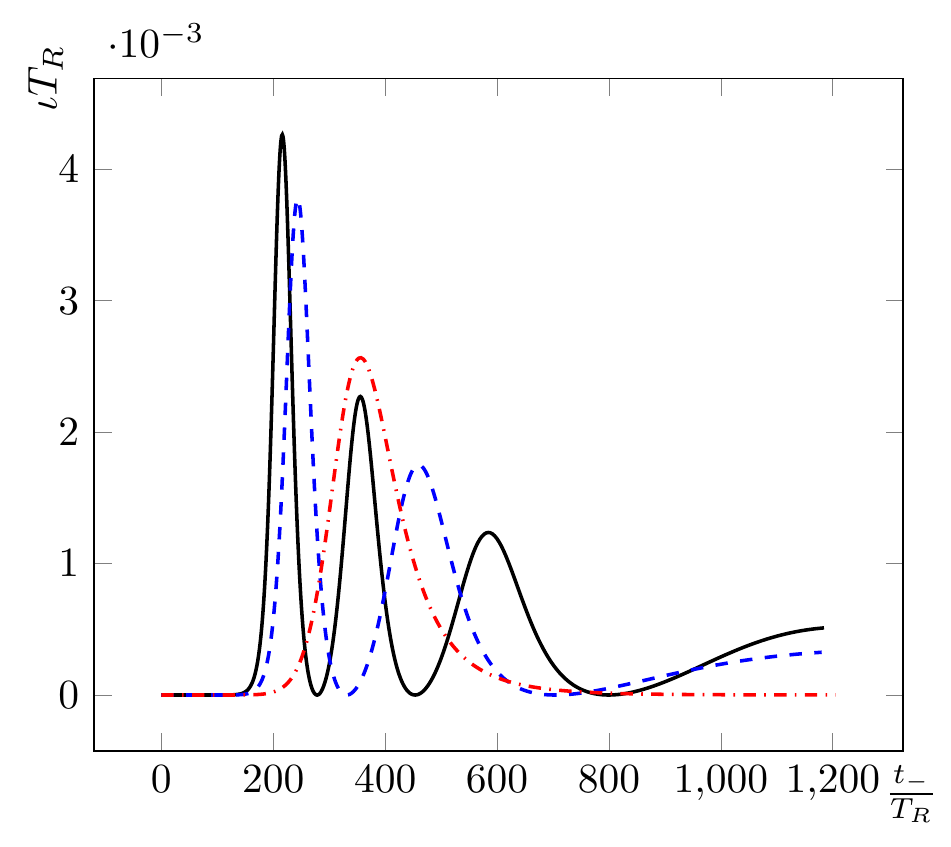}
     \caption{Radiance $\iota T_R$ of a non-ideal sample of positronium atoms as a function of dimensionless time $t_-/T_R$. The number of atoms is maximal and corresponds to (\ref{eq138}), but dephasing effects with a common time scale $T_{dph}$ are present. The solid curve corresponds to $T_{dph}=400\,T_R$, the dashed blue curve corresponds to $T_{dph}=300\,T_R$, and the dashed red curve corresponds to $T_{dph}=200\,T_R$.}
\label{Fig2}
\end{figure}

For simplicity, let us assume $T_1=T_2=T_{dph}$ and introduce the dimensionless dephasing time $\tau_{dph}=T_{dph}\,\Omega$ together with other dimensionless variables (\ref{eq91}). Then instead of (\ref{eq104}) we get
\begin{equation} 
\frac{\partial R}{\partial \tau_-}=Z\epsilon-\frac{R}{\tau_{dph}},\;\;\;
\frac{\partial Z}{\partial \tau_-}=-R\epsilon-\frac{Z}{\tau_{dph}},\;\;\;
\frac{\partial \epsilon}{\partial \xi}=R\,.
\label{eq149}
\end{equation} 
The first two equations show that $\frac{1}{2}\frac{\partial}{\partial \tau_-}\left(R^2+Z^2\right)=-\frac{R^2+Z^2}{\tau_{dph}}$ and, therefore, $R^2+Z^2=B^2e^{-2\tau_-/\tau_{dph}}$, where $B$ is some constant (in fact, all that follows from the above is that $B$ does not depend on $\tau_-$, but may depend on $\xi$). Therefore, we can again introduce the Bloch angle $\theta(\tau_-,\xi)$ through
\begin{equation}
    R=Be^{-\frac{\tau_-}{\tau_{dph}}}\sin{\theta},\;\;\;
    Z=Be^{-\frac{\tau_-}{\tau_{dph}}}\cos{\theta}.
 \label{eq150}   
\end{equation}
Then either from the first, or the second equation of (\ref{eq149}) we obtain $\epsilon=\frac{\partial \theta}{\partial \tau_-}$, while the last equation will give
\begin{equation}
\frac{\partial^2 \theta}{\partial \xi \partial \tau_-}= Be^{-\frac{\tau_-}{\tau_{dph}}}\sin{\theta}.
\label{eq151}
\end{equation}
Again we can consider solutions that depend on only one variable \cite{Rajabi_2016-I}
\begin{equation}
 q=2\sqrt{\xi\tau_{dph}\left(1-e^{-\frac{\tau_-}{\tau_{dph}}}\right)}.
\label{eq152}
\end{equation}
Then
\begin{equation}
    \frac{\partial}{\partial \tau_-}=\sqrt{\frac{\xi}{\tau_{dph}\left(1-e^{-\frac{\tau_-}{\tau_{dph}}}\right)}}\;e^{-\frac{\tau_-}{\tau_{dph}}}\;\frac{d}{dq},\;\; \frac{\partial}{\partial \xi}=\sqrt{\frac{\tau_{dph}\left(1-e^{-\frac{\tau_-}{\tau_{dph}}}\right)}{\xi}}\;\frac{d}{dq},
\label{eq153}    
\end{equation}
and from (\ref{eq152}) we again obtain the Burnham-Chiao equation
\begin{equation}
 \frac{d^2\theta(q)}{d q^2}+\frac{1}{q}\,\frac{d\theta(q)}{dq}=B\sin{\theta(q)},\;\;q=2\sqrt{\frac{z}{L}\,\frac{T_{dph}}{T_{R}}\,\left(1-e^{-t_-/T_{dph}}\right)}\,,
 \label{eq154}
\end{equation}
the numerical solution of which can be used to calculate the dimensionless radiance $\iota T_R$ again according to 
\begin{equation}
\iota T_R=\frac{1}{4}\,\frac{T_R}{t_-}\left(\frac{d\theta}{dq}\right)^2.
\label{eq155}
\end{equation}
However, now
\begin{equation}
 \frac{t_-}{T_R}=-\frac{T_{dph}}{T_R}\,\ln{\left(1-\frac{q^2}{4}\,\frac{T_R}{T_{dph}}\right)},
 \label{eq156}
\end{equation}
and, unlike the ideal case, the range of $q$ is finite and limited by the interval $[0,2\sqrt{T_{dph}/T_ R}]$. Therefore, if the upper bound on $q$ is less than $q_D$, the superradiance burst will never occur. This leads to the necessary condition $T_{dph}>T_D$ for the occurrence of superradiance (more general condition when $T_1\ne T_2$ is discussed in \cite{Rajabi_2020}). However, even if this condition is satisfied, but $T_{dph}$ remains of the order of $T_D$, the superradiance bursts will weaken and the ringing may disappear completely. This is shown in Fig. \ref{Fig2} for a positronium sample for which $T_D\approx 150\,T_R$.
The non-ideal case may also include the situation where the Arecchi-Courtens condition takes the form $L=\eta\, c\,T_R$ with $\eta<1$, i.e. the number of atoms participating in the cooperative behavior is less than the maximum number given by (\ref{eq138}). Using (\ref{eq102}) for $T_{SR}=T_R/2$ in the condition $L=\eta\, c\,T_R$ and assuming a unit Fresnel number, we obtain for positronium sample, in the form of a long cylinder of radius $r$ and length $L$, $N=\frac{4}{3}\eta\omega\tau_0=\eta N_{max}$ and
\begin{eqnarray} &&
 L=\sqrt{\frac{8\pi\tau_0\nu^2\eta}{3cn}}\approx 1.86\cdot 10^{10}\sqrt{\frac{\eta}{n}}~\mathrm{cm},\nonumber \\ && 
 r=\left(\frac{8c\tau_0\eta}{\pi n}\right )^{1/4}\approx 
 3.89\cdot 10^4\left(\frac{\eta}{n}\right)^{1/4}~\mathrm{cm}, \nonumber \\ && T_R=\sqrt{\frac{8\pi\tau_0\nu^2}{3c^3n\eta}}\approx\frac{0.62}{\sqrt{\eta n}}~\mathrm{sec}.
 \label{eq157}
\end{eqnarray}
In these formulas, the number of positronium atoms per unit volume $n$ must be expressed in units of $\mathrm{cm}^{-3}$. For example, for $\eta=10^{-3}$ and $n=10~\mathrm{cm}^{-3}$, we have  $L\approx 1.9\cdot 10^3~\mathrm{km}$, $r\approx 39~\mathrm{m}$, $T_R\approx 6.2~\mathrm{sec}$ and $T_D\approx 11.4~\mathrm{min}$.  In an astrophysical context, these numbers do not seem unrealistic. An estimate for $T_D$ was obtained by numerically solving (\ref{eq143}) for $\theta_0=2/\sqrt{N}$, which yields $T_D\approx 150\,T_R$, $T_D\approx 136\,T_R$, $T_D\approx 123\,T_R$, and $T_D\approx 110\,T_R$ for $\eta=1$, $\eta=0.1$, $\eta=0.01$, and $\eta=0.001$, respectively.

In the case of hydrogen without dephasing effects, the analogues results are
\begin{equation}
 L\approx 4.4\cdot 10^{11}\sqrt{\frac{\eta}{n}}~\mathrm{cm},\;\;\; 
 r\approx 2.3\cdot 10^6\left(\frac{\eta}{n}\right)^{1/4}~\mathrm{cm}, \;\;\; T_R\approx\frac{14.8}{\sqrt{\eta n}}~\mathrm{sec},
 \label{eq158}
\end{equation}
and $T_D\approx 229\,T_R$, $T_D\approx 212\,T_R$, $T_D\approx 195\,T_R$ and $T_D\approx 179\,T_R$ for $\eta=1$, $\eta=0.1$, $\eta=0.01$, and $\eta=0.001$, respectively. For example, for $\eta=0.05$ and $n=1~\mathrm{cm}^{-3}$, we have  $L\approx 1.0\cdot 10^{6}~\mathrm{km}$, $r\approx 1.1\cdot 10^4~\mathrm{m}$, $T_R\approx 66.2~\mathrm{sec}$ and $T_D\approx 3.8~\mathrm{hr}$ (for $\eta=0.05$ the numerical solution of (\ref{eq143}) gives $T_D\approx 206.6\,T_R$).

Doppler broadening in thermally relaxed regions of the interstellar medium is the main dephasing mechanism that makes the development of superradiance unlikely, since the corresponding dephasing time is very short (about $10^{-3}~\mathrm{sec}$ at $T=100~\mathrm{K}$) \cite{Rajabi_2016-I}. For superradiance to occur, three conditions must be met: a population inversion, sufficient velocity coherence between the atoms involved in the effect to overcome Doppler broadening, and any other dephasing effects present must occur on time scales longer than those that characterize superradiance. Thus, we do not expect superradiance from thermally relaxed regions, but from regions where a significant amount of energy is suddenly released, creating a flow of atoms in strong thermal disequilibrium but with strong velocity coherence \cite{Rajabi_2016-I}. The ubiquitous presence of cosmic masers, which also require population inversion and velocity coherence, suggests that the first two conditions are not so exotic in space. However, for superradiance to occur, a third condition must also be met.

The problem of Doppler broadening can be briefly described as follows: molecules with excessive velocity differences do not see a common radiation field corresponding to their natural Doppler-shifted frequencies and, therefore, cannot evolve into the highly entangled state required for superradiance. It is in this sense that Doppler broadening can be considered as a dephasing mechanism of central importance in astrophysical applications of superradiance, since under astrophysical conditions the total bandwidth of the velocity distribution is typically several orders of magnitude wider than the bandwidth of the superradiance transient. However, a wide velocity distribution does not prevent superradiance in regions with thermal disequilibrium, provided that the inverted column density in at least one velocity slice still exceeds the superradiance threshold. A detailed discussion of the effects of astrophysical velocity distributions upon superradiance can be found in \cite{Wyenberg_2022}.

The dephasing time scale $T_{dph}$ is difficult to estimate without detailed knowledge of the environmental conditions in the region with population inversion and velocity coherence. In the case of atomic hydrogen, elastic collisions between hydrogen atoms are considered to give the shortest time $T_{dph}$ \cite{Rajabi_2016-I}. Consequently, we estimate
$T_{dph}=\frac{1}{n\sigma v}$, where $\sigma$ is the elastic scattering cross section and $v$ is the average impact velocity. For $\sigma$ we take an old \cite{Purcell_1956} semi-classical low-energy estimate $\sigma\approx 2\cdot 10^{-14}~\mathrm{cm}^2$. The average impact velocity, of course, depends on the degree of velocity coherence. Somewhat conservatively, we assume that $v\approx 1.3\cdot 10^5~\mathrm{cm}/\mathrm{sec}$, which corresponds to an effective temperature of $T=100~\mathrm{K}$. Then $T_{dph}\approx \frac{3.8\cdot 10^8}{n}~\mathrm{sec}$. From (\ref{eq158}), expressing $\sqrt{\eta}$ in terms of $L$ and substituting into $T_R$, we obtain $T_R\approx \frac{6.5\cdot 10^{12}}{Ln}~\mathrm{sec}$, and since $T_D$ according to our calculations is less than approximately $230\,T_R$, the condition $T_D<T_{dph}$ will give $L>3\cdot 10^6~\mathrm{cm}$. On the other hand, since $\eta<1$, it follows from (\ref{eq158}) that $L<\frac{4.4\cdot 10^{11}}{\sqrt{n}}~\mathrm{cm}$. Therefore, the concentration of hydrogen atoms $n$ cannot be too large, namely $n<2\cdot 10^{10}~\mathrm{cm}^{-3}$, and, consequently, $T_R$ cannot be too short: $T_R>10^{-4}~\mathrm{sec}$. 

Let us summarize our rough estimates in the case of hydrogen:
\begin{equation}
3\cdot 10^6~\mathrm{cm}<L<\frac{4.4\cdot 10^{11}}{\sqrt{n}}~\mathrm{cm},\;\;
n<2\cdot 10^{10}~\mathrm{cm}^{-3},\;\;T_R>10^{-4}~\mathrm{sec}.
\label{eq159}
\end{equation}

In the case of positronium, in view of the expected abundance of electrons and positrons at the production site, we assume that the main dephasing mechanism is scattering of electrons and positrons on positronium, and use the zero-energy scattering cross section of electrons and positronium estimated in \cite{Mitchell_2025} to be approximately $210\,\pi a_B^2\approx 1.85\cdot 10^{-14}~\mathrm{cm}^2$, which is almost equal to the elastic scattering cross section of hydrogen used above. Since most positron annihilation near the Galactic Center occurs in the warm neutral and warm ionized phases of the interstellar medium, where the temperature is approximately $8000~\mathrm{K}$, we take $V=5\cdot 10^7$ cm/s as a rough estimate of the average electron impact velocity, which corresponds to this temperature. As a result we obtain $T_{dph}\approx \frac{10^6}{n}~\mathrm{sec}$, and using (\ref{eq157}) and $T_D<150\,T_R$, we get for positronium
\begin{equation}
1.7\cdot 10^6~\mathrm{cm}<L<\frac{1.86\cdot 10^{10}}{\sqrt{n}}~\mathrm{cm},\;\;
n<10^8~\mathrm{cm}^{-3},\;\;T_R>6\cdot 10^{-5}~\mathrm{sec}.
\label{eq160}
\end{equation}

\section{Concluding remarks}
The first suggestion of the existence of a bound state of an electron and a positron, made by Mohorovicic\v{c}i\'{c} \cite{Mohorovicic_1934}, assumed that it would be discovered in the course of astrophysical observations. However, in fact it was discovered in the laboratory, and experimental study of its properties has advanced considerably since its discovery \cite{Cassidy_2018}, while the search for astrophysical manifestations of positronium continues to this day.

Burdyuzha et al. suggested \cite{Burdyuzha_1997,burdyuzha_1999} that the positronium spin-flip line, enhanced by the maser effect (idea pioneered by Shklovskii \cite{Shklovskii_1967}), could be detected by next-generation radio telescopes. Despite the superficial similarity, the maser effect is fundamentally different from Dicke superradiance. The maser effect is a collective but not coherent phenomenon \cite{Rajabi_2016-I}. It can be decomposed down into successive individual acts of stimulated emission of photons by the incident radiation, as a result of which the number of photons in the masing sample grows exponentially until it is saturated and reaches a steady state, when some equilibrium is established in the population inversion, depending on the pumping intensity. The maser can operate continuously as long as the pump source is on.

In contrast, in Dicke superradiance, either several atoms form a quantum state and emit a powerful burst of radiation as a single quantum system (small-size superradiance), or several semiclassical dipoles become phase-locked due to interaction with the common radiation field and again emit a superradiance pulse as a single larger dipole (large-size superradiance). In this case, the radiation pulse cannot be decomposed into more primitive successive events (but the ringing phenomenon can occur in the superradiance of a large sample if the dephasing effects are insignificant).

Our results (\ref{eq159}) and (\ref{eq160}) indicate that Dicke superradiance is possible for the spin-flip lines of hydrogen and positronium under astrophysical conditions (and for hydrogen we confirm the findings of \cite{Rajabi_2016-I}, the pioneering paper that first considered Dicke superradiance in an astrophysical context).

Of course, there is a fundamental difference between hydrogen and positronium: the former is stable, while the latter annihilates very quickly. However, we assume that the continuous process of positronium production helps to maintain a nearly constant positronium density for a sufficiently long time. Nevertheless, this aspect certainly requires more attention than is given in this article. Note that in \cite{Vlasov_1989,Cui_2012} it was proposed to manipulate the dynamics of annihilation of a dense gas of positronium atoms under laboratory conditions using the superradiance and subradiance modes of cooperative spontaneous emission of the system, and in \cite{Cui_2012} the existence of an annihilation channel was taken into account by introducing a third (vacuum) state.

There is another important difference between the 21 cm hydrogen line and the 203 GHz positronium line as far as the population inversion is concerned: for hydrogen, a special mechanism is needed to generate the population inversion (hydrogen atoms can change their hyperfine state by absorbing and re-emitting $Ly\alpha$ photons, the so-called Wouthuysen-Field effect \cite{Dijkstra_2006}), whereas for positronium, the population inversion is expected to be generated automatically due to the huge difference in the lifetimes (in vacuum) of ortho- and para-positronium, and also due to the threefold higher statistical weight of ortho-positronium in positronium formation processes. Of course, in a medium, the ortho-positronium lifetime can be significantly shorter (the so-called ortho-positronium quenching \cite{Hyodo_2009}), and in an astrophysical context, collisions of positronium with electrons and positrons will reduce the magnitude of the population inversion. However, we expect that the quenching is controlled by the value of $T_{dph}$ and will not be very significant if the condition $T_{dph}>T_D$ is satisfied (especially if $T_{dph}\gg T_D$). 

Interestingly, it may be that hydrogen spin-flip line superradiance from an astrophysical source has already been observed under rather unusual circumstances \cite{mendez_2024}. The former Ohio State University radio observatory, known as the Big Ear, was part of the university's Search for Extraterrestrial Intelligence (SETI) project: a systematic search for narrow-band radio signals around the 1420 MHz hydrogen line. On August 15, 1977, they detected a short but intense narrow-band radio signal exhibiting the characteristics expected of a technological signal emanating from space. The signal, nicknamed ``Wow!" because of astronomer Jerry Ehman's ``Wow!" annotation on the data printout, caused a great stir in the SETI community \cite{mendez_2025}. The source of the ``Wow!" signal has never been found, despite numerous attempts to detect it, and remains a mystery to this day.

It has recently been suggested that Dicke superradiance is the most plausible mechanism to explain the Wow! signal \cite{mendez_2024}. Remarkably, the hypothesis explains all the main features of the Wow! signal, namely its narrow-band frequency, duration, intensity, lack of modulation, and rarity. Thus, the Wow! signal may represent the first documented case of astrophysical superradiance of the 21 cm atomic hydrogen line. A magnetar flare or soft gamma repeater could serve as a pump source to create a population inversion in a cylindrical sample of hydrogen atoms of sufficient length. Note that observations of some hydrogen clouds have revealed a network of thin, elongated filaments of cold atomic hydrogen, with individual strands reaching $17~\mathrm{pc}\approx 5.2\cdot 10^{19}~\mathrm{cm}$ in length and only $0.1~\mathrm{pc}\approx 3.1\cdot 10^{17}~\mathrm{cm}$ in width \cite{mendez_2024} (however, for 21 cm line such a geometry corresponds to a very large Fresnel number).

It is quite clear that the search of Wow! like signals to confirm or disprove their superradiance nature is actual and interesting problem in radio astronomy. Current observational strategies typically focus on broadband signals and long integration times \cite{mendez_2024}, which probably explains why the hydrogen spin-flip line superradiance has so far eluded reliable detection: such a signal could be a very short-lived burst of emission, namely (\ref{eq159}) indicates that the lower bound of $T_R$ is less than millisecond. The same is true for the positronium spin-flip line superradiance, which also represents a very interesting observational problem in millimeter-wave radio astronomy. We hope that this paper will help to attract the attention of the astrophysical community to the exciting possibility of astrophysical superradiance and stimulate both theoretical and observational activity in this new emerging field.

\appendix
\section{Dicke superradiance: a classical analogue}
There are interesting and elegant classical analogues of Dicke superradiance \cite{Dicke_1954,Eberly_1971,Menshikov_1998}. In particular, let us consider $N$ identical magnets with magnetic moment $\mu$ each, precessing in a constant external magnetic field $\bf{B}$ with angular velocity $\boldsymbol{\omega}=-g\mathbf{B}$ under the action of the radiation reaction, where $g$ is the gyromagnetic ratio \cite{Eberly_1971,Menshikov_1999}. It is assumed that the polar angle $\theta(t)$ is the same for all magnets and does not change with time, but the azimuth angles $(\varphi_i-\omega t)$ change with time and their initial values $\varphi_i$ may be different.

\begin{figure}[ht]
    \centering
    \includegraphics[width=1\textwidth]{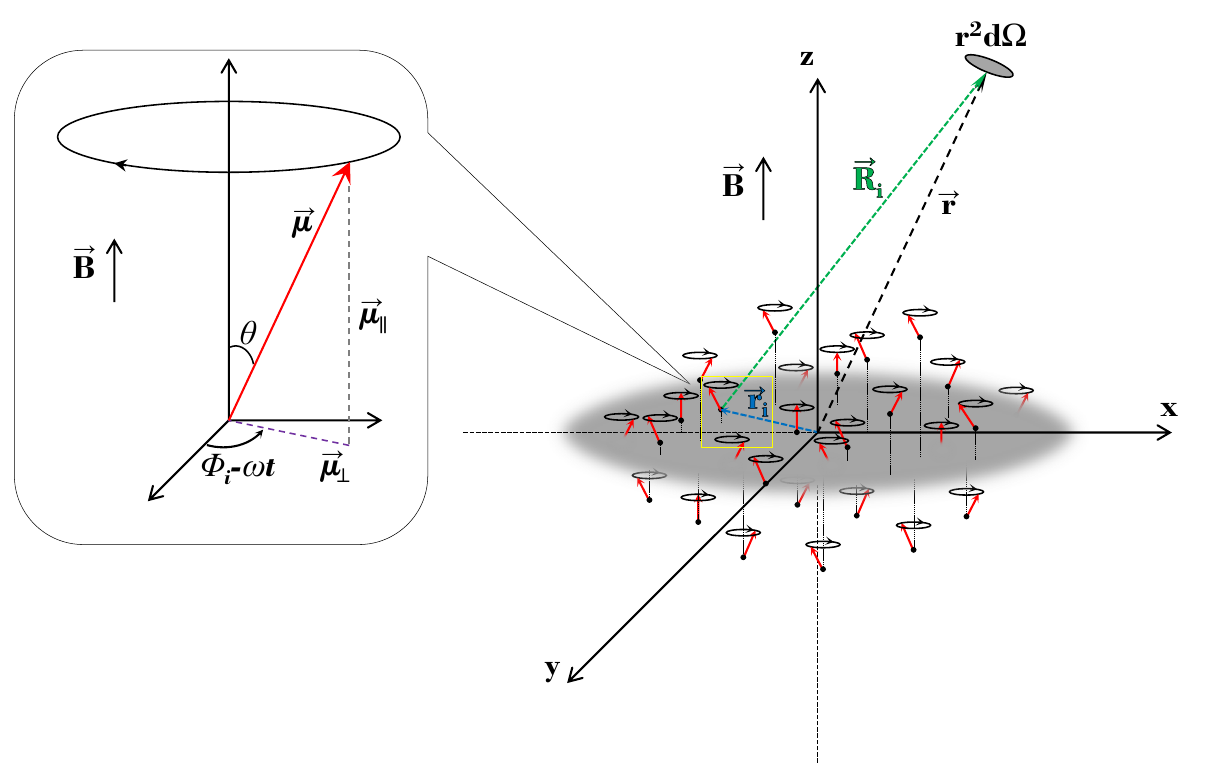}
     \caption{An illustration of a classical analogue in which Dicke superradiance can occur: $N$ identical magnets with magnetic moment $\mu$ each, precessing in a constant external magnetic field $\bf{B}$ with angular velocity $\boldsymbol{\omega}$ under the action of the radiation reaction. The inset shows the configuration of a typical magnet precessing in the magnetic field $\bf{B}$. The scale of $r_i$ as appears in the figure is used only for the purpose of a clear illustration. Both the gray horizontal circle in the $(x,y)$ plane and the vertical dotted lines between it and the magnets are imaginary objects; they were added only to help picturing the group of magnetic dipoles in the $3D$ space around the origin.}
\label{classical}
\end{figure}

Fig. \ref{classical} illustrates the system described and shows the configuration of a typical magnet precessing in the magnetic field $\bf{B}$. In our problem, we will have $r_i\ll r$, and therefore the scale of $r_i$ as appears in the figure is used only for the purpose of a clear illustration. Also, both the gray horizontal circle in the $(x,y)$ plane and the vertical dotted lines between it and the magnets are imaginary objects; they were added only to help picturing the group of magnetic dipoles in the $3D$ space around the origin.

In case of one oscillating dipole, the intensity of the electromagnetic radiation per solid angle $d\Omega$ at the observation point having the radius-vector ${\bf{R}}$ equals \cite{Landau_2010}
\begin{equation}
    \frac{dI}{d\Omega}=\frac{cH^2}{4\pi}R^2,
    \label{eq2}
\end{equation}
where $H$ is the magnetic field created by the dipole at the observation point and $c$ is the speed of light. \cite{Landau_2010}
\begin{equation}
    {\bf{H}}=\frac{1}{c^2 R}\left (\ddot{\boldsymbol{\mu}}\times{\mathbf{n}}\right )\times{\mathbf{n}}.
\label{eq3}
\end{equation}
In our case, $\boldsymbol{\mu}=\boldsymbol{\mu}_\parallel+\boldsymbol{\mu}_\perp$, where $\boldsymbol{\mu}_\parallel=N\mu\cos{\theta}\,\mathbf{k}$ is a constant and does not participate in the radiation, while at retarded time
\begin{equation}
 \boldsymbol{\mu}_\perp\left(t-\frac{R_i}{c}\right)=\mu\sin{\theta}\sum\limits_{i=1}^N \left \{\cos{\left[\omega\left(t-\frac{R_i}{c}\right)-\varphi_i\right ]}\,\mathbf{i}-\sin{\left[\omega\left(t-\frac{R_i}{c}\right)-\varphi_i\right ]}\,\mathbf{j}\right\}  
 \label{eq4}
\end{equation}

Assuming that the radius vector of the i-th dipole is $\mathbf{r}_i$ much smaller than the distance between observation point and the origin $\mathbf{r}$, $r_i\ll r$, so that  $$R_i=\sqrt{r^2-2\mathbf{r}\cdot\mathbf{r}_i+r_i^2}\approx r-\mathbf{n}\cdot\mathbf{r}_i,$$ where $\mathbf{n}=\mathbf{r}/r$, we will have 
\begin{equation}
 \boldsymbol{\mu}_\perp=\mu\sin{\theta}\left\{\left (\sum\limits_{i=1}^N \cos{\Phi_i}\right)\,\mathbf{i}-\left(\sum\limits_{i=1}^N\sin{\Phi_i}\right)\,\mathbf{j}\right\},\;\;\Phi_i= \omega\left(t-\frac{r}{c}\right)+\mathbf{k}\cdot\mathbf{r}_i-\varphi_i,
 \label{eq5}
\end{equation}
where $\mathbf{k}=\frac{\omega}{c}\mathbf{n}$ is the wave vector of the radiation field.
Therefore
\begin{equation}
    \frac{dI}{d\Omega}=\frac{1}{4\pi c^3}|\ddot{\boldsymbol{\mu}}\times\mathbf{n}|^2=\frac{\omega^4}{4\pi c^3}\left [\mu_\perp^2-(\boldsymbol{\mu}_\perp\cdot\mathbf{n})^2\right].
    \label{eq6}
\end{equation}
Now we average over the fast Larmor precession with frequency $\omega$, noting that when using the trigonometric product-to-sum formulas and 
$$\overline{\cos{(\Phi_i+\Phi_j)}}=0,\;\;\overline{\sin{(\Phi_i+\Phi_j)}}=0$$ 
we obtain
\begin{eqnarray} &&
\overline{\left(\sum\limits_{i=1}^N\cos{\Phi_i}\right)^2}=\overline{\sum\limits_{i,j=1}^N \cos{\Phi_i}  \cos{\Phi_j}}=\frac{1}{2}\sum\limits_{i,j=1}^N\cos{(\Phi_i-\Phi_j)},\nonumber \\ && \overline{\left(\sum\limits_{i=1}^N\sin{\Phi_i}\right)^2}=\overline{\sum\limits_{i,j=1}^N \sin{\Phi_i}  \sin{\Phi_j}}=\frac{1}{2}\sum\limits_{i,j=1}^N\cos{(\Phi_i-\Phi_j)}, \\ &&
\overline{\sum\limits_{i,j=1}^N\sin{\Phi_i}\cos{\Phi_j}}=\frac{1}{2}\sum\limits_{i,j=1}^N\sin{(\Phi_i-\Phi_j)}=0.\nonumber
\label{eq7}
\end{eqnarray}
As a result, we get
\begin{equation}
\frac{d\bar{I}}{d\Omega}= \frac{\omega^4\mu^2}{8\pi c^3}\left(2-n_x^2-n_y^2\right)\sin^2{\theta} \sum\limits_{i,j=1}^N\cos{(\Phi_i-\Phi_j)}.  
\label{eq8}
\end{equation}
But
\begin{equation}
\sum\limits_{i,j=1}^N\cos{(\Phi_i-\Phi_j)}=\sum\limits_{i,j=1}^N\cos{[(\mathbf{k}\cdot\mathbf{r}_i-\varphi_i)-(\mathbf{k}\cdot\mathbf{r}_j-\varphi_j)]}=\left|\sum\limits_{i=1}^Ne^{i(\mathbf{k}\cdot\mathbf{r}_i-\varphi_i)}\right |^2,
\label{eq9}
\end{equation}
and finally \cite{Eberly_1971}
\begin{equation}
\frac{d\bar{I}}{d\Omega}=N^2I_0(\mathbf{n})\sin^2{\theta}\left|\frac{1}{N}\sum\limits_{i=1}^Ne^{i(\mathbf{k}\cdot\mathbf{r}_i-\varphi_i)}\right |^2,
\label{eq10}
\end{equation}
where
\begin{equation}
I_0(\mathbf{n})=\frac{\omega^4\mu^2}{8\pi c^3}\left(1+n_z^2\right).
\label{eq11}
\end{equation}
Introducing the dimensionless electromagnetic energy stored in the system of magnetic dipoles $W=\frac{-N\boldsymbol{\mu}\cdot\mathbf{B}}{2\mu B}=-\frac{N}{2}\cos{\theta}$ and equating its time derivative (energy loss per unit time) to the total radiated power $\bar{I}$, we obtain
\begin{equation}
-\frac{dW}{dt}=\frac{2}{\mu B}\left(\frac{N}{2}-W\right)\left(\frac{N}{2}+W\right)\int d\Omega I_0(\mathbf{n}) \left|\frac{1}{N}\sum\limits_{i=1}^Ne^{i(\mathbf{k}\cdot\mathbf{r}_i-\varphi_i)}\right |^2.
\label{eq12}
\end{equation}
It is convenient to introduce the constant $\tau_0$ of dimension of time through
\begin{equation}
\frac{1}{\tau_0}=\frac{2}{\mu B}\int d\Omega I_0(\mathbf{n})=\frac{4\omega^4\mu}{3Bc^3},
\label{eq13}
\end{equation}
and rewrite (\ref{eq12}) in the form
\begin{equation}
-\frac{dW}{dt}=\frac{\lambda}{\tau_0}\left(\frac{N}{2}-W\right)\left(\frac{N}{2}+W\right),   
\label{eq14}
\end{equation}
where \cite{Eberly_1971}
\begin{equation}
\lambda=\frac{2\tau_0}{\mu B}\int d\Omega I_0(\mathbf{n}) \left|\frac{1}{N}\sum\limits_{i=1}^Ne^{i(\mathbf{k}\cdot\mathbf{r}_i-\varphi_i)}\right |^2.  
\label{eq15}
\end{equation}
The dimensionless parameter $\lambda$ is defined such that $\lambda=1$ if the sample size is much smaller than the radiation wavelength, so that $\mathbf{k}\cdot\mathbf{r}_i\approx 0$ for all $i$, and all magnets precess in phase ($\varphi_i=\varphi_j$ for all $i,j$).

The differential equation (\ref{eq14}) is easily solved by separation of variables, resulting in
\begin{equation}
W(t)=-\frac{N}{2}\tanh{\left [\frac{N\lambda}{2\tau_0}(t-t_0)\right]}.
\label{eq16}
\end{equation}
Therefore, the radiation intensity is
\begin{equation}
I(t)=-\frac{d}{dt} (2\mu B\,W)  =\frac{N^2\lambda}{2\tau_0}\,\mu B\,\sech^2{\left [\frac{N\lambda}{2\tau_0}(t-t_0)\right]} 
\label{eq17}
\end{equation}
As can be seen from (\ref{eq17}), the radiation intensity is not monotonic: after a delay time $t_0$ there is an intensity peak whose duration $\tau=\frac{\tau_0}{N\lambda}$ scales as $1/N$ (where $\tau_0$ is the characteristic time of spontaneous decay). These are characteristic features of Dicke superradiance, as well as the proportionality of the peak intensity to $N^2$. Inspired by the work of Dicke \cite{Dicke_1954}, Bloom \cite{Bloom_1956} derived equations similar to (\ref{eq16}) and (\ref{eq17}) using semiclassical radiation theory to describe the reaction of a collection of two-state molecules driven by an external electromagnetic field.

The argument used to derive the equation (\ref{eq12}) is based solely on conservation of energy and ignores the effect of the radiation reaction on the precession phase \cite{Crisp_1989}. The correct equation of motion of a point magnetic dipole precessing in a magnetic field, which includes the radiation reaction, is \cite{Ginzburg_1943,Ginzburg_2013,Crisp_1989}
\begin{equation}
\frac{d}{dt}\left(\boldsymbol{\mu}+\alpha\boldsymbol{\mu}\times\dot{\boldsymbol{\mu}}\right )=
g\,\boldsymbol{\mu}\times\mathbf{B}+\frac{2g}{3c^3}\boldsymbol{\mu}\times\dddot{\boldsymbol{\mu}},
\label{eq18}
\end{equation}
where $g$ is the gyromagnetic ratio and $\alpha$ is some constant. The term $\alpha\boldsymbol{\mu}\times\dot{\boldsymbol{\mu}}$ corresponds to the angular momentum of electromagnetic origin, analogous to the electromagnetic Lorentz mass $\frac{e^2}{2ac^2}$ (for an electron as a uniformly charged sphere of radius $a$) \cite{Rohlich_1960,Schwinger_1983}. For a highly conducting magnetized sphere of radius $a$, one can find \cite{Ginzburg_2013} that $\alpha=\frac{2}{3ac^2}$ and it diverges in the point-like limit $a\to 0$, just like the electromagnetic mass.

Without the radiation reaction $\dot{\boldsymbol{\mu}}=g\,\dot{\boldsymbol{\mu}}\times\mathbf{B}$ and, therefore,
$$\mu_z^{(0)}=\mu\cos{\theta}=\mathrm{const},\;\; \boldsymbol{\mu}^{(0)}_\perp=\mu\sin{\theta}\left(\cos{(\omega t)}\,\mathbf{i}- \sin{(\omega t)}\,\mathbf{j}\right).$$
Assuming that the terms associated with the radiation reaction are small, we can replace $\ddot{\boldsymbol{\mu}}_\perp$ in them by $-\omega^2\,\boldsymbol{\mu}_\perp$, which holds for $\boldsymbol{\mu}_\perp^{(0)}$, and from (\ref{eq18}) we obtain
\begin{equation}
\dot{\boldsymbol{\mu}}=\alpha\omega^2\boldsymbol{\mu}_\parallel\times\boldsymbol{\mu}_\perp+
g\,\boldsymbol{\mu}\times\mathbf{B}-\frac{2g\omega^2}{3c^3}\boldsymbol{\mu}\times\dot{\boldsymbol{\mu}}_\perp.
\label{eq19}
\end{equation}
From (\ref{eq19}) it follows that the longitudinal component of the vector $\dot{\boldsymbol{\mu}}$ satisfies the equation
\begin{equation}
\dot{\mu}_z=-\frac{2g\omega^2}{3c^3}\left(\boldsymbol{\mu}_\perp\times\dot{\boldsymbol{\mu}}_\perp\right)\cdot\mathbf{k}\approx -\frac{2g\omega^2}{3c^3}\left(\boldsymbol{\mu}_\perp^{(0)}\times\dot{\boldsymbol{\mu}}_\perp^{(0)}\right)\cdot\mathbf{k},
\label{eq20}
\end{equation}
which is the same as 
\begin{equation}
\dot{\mu}_z=\frac{2g\omega^3\mu^2}{3c^3}\sin^2{\theta}=\frac{2\omega^4\mu^2}{3Bc^3}\left(1+\frac{\mu_z}{\mu}\right)\left(1-\frac{\mu_z}{\mu}\right),
\label{eq21}
\end{equation}
and is equivalent to (\ref{eq14}) with $N=1$.

For the transverse components of $\boldsymbol{\mu}$ from (\ref{eq19}) we get 
\begin{equation}
\dot{\mu}_x=\left(\omega-\alpha\omega^2\mu_z\right)\mu_y+\frac{2\omega^3\mu_z}{3Bc^3}\,\dot{\mu}_y,\;\;\;
\dot{\mu}_y=-\left(\omega-\alpha\omega^2\mu_z\right)\mu_x-\frac{2\omega^3\mu_z}{3Bc^3}\,\dot{\mu}_x.
\label{eq22}
\end{equation}
By introducing $Z=\mu_x+i\mu_y$ and keeping only the first-order terms for small parameters, (\ref{eq22}) can be rewritten as follows
\begin{equation}
    \frac{\dot{Z}}{Z}=-i\omega-\frac{2\omega^4}{3Bc^3}\left(1+i\,\frac{3\alpha Bc^3}{2\omega^2}\right)\mu_z,
\label{eq23}
\end{equation}
which can be integrated by using
\begin{equation}
\mu_z=\mu\tanh{\left(\frac{2\omega^4\mu}{3Bc^3}\,t\right)},\;\;\int\limits_0^t\mu_z(\tau)d\tau=\frac{3Bc^3}{2\omega^4}\ln{\left[\cosh{\left(\frac{2\omega^4\mu}{3Bc^3}\,t\right)}\right]},   
\label{eq24}
\end{equation}
which follow from (\ref{eq22}). As a result, we obtain \cite{Crisp_1989}
\begin{equation}
 \mu_x+i\mu_y=\mu e^{-i\omega t}\left [ \sech{\left(\frac{2\omega^4\mu}{3Bc^3}\,t\right)}\right]^{1+i\,\frac{3\alpha Bc^3}{2\omega^2}}.
 \label{eq25}
\end{equation}
According to (\ref{eq22}), (\ref{eq25}) can be approximately interpreted as indicating a small time-dependent shift of the Larmor frequency \cite{Crisp_1989}
\begin{equation}
 \Delta \omega(t)=-\alpha\omega^2\mu_z(t)= -\alpha\omega^2\mu\,\tanh{\left(\frac{2\omega^4\mu}{3Bc^3}\,t\right)}. 
 \label{eq26}
\end{equation}

If initially all magnets are inverted so that $W(0)=\frac{N}{2}$, then (\ref{eq16}) indicates that $t_0=\infty$, i.e. the system never radiates. This is not surprising, since such a classical system is in an unstable equilibrium. Although many aspects of Dicke superradiance can be understood using semiclassical concepts, its initial phase requires quantum concepts and formalism, as do many other purely quantum mechanical aspects of superradiance \cite{Eberly_1972,Zheleznyakov_1989,Gross_1982}.

\section{Multipole Hamiltonian of the light-matter interactions}
There are two widely used ways of describing the interaction of a nonrelativistic quantum mechanical system consisting of charged particles with masses $m_\alpha$ and charges $q_\alpha$ with an electromagnetic radiation field with vector potential $\mathbf{A}(\mathbf{r},t)$ when the characteristic wavelengths of the radiation are much larger than the dimensions of the quantum system. In the minimal coupling formulation, the canonical momenta $\hat{\mathbf{p}}$ in the field-free Hamiltonian are replaced by $\hat{\mathbf{p}}-\frac{q}{c}\mathbf{A}$, which leads to the Hamiltonian (for spinless particles)
\begin{equation}
    \hat H^{(A)}=\sum\limits_\alpha\frac{1}{2m_\alpha}\left (\hat{\mathbf{p}}_\alpha-\frac{q_\alpha}{c}\mathbf{A}({\bf{r}}_\alpha,t)\right )^2+V_c+\sum\limits_{\mathbf{k},\sigma}\hbar\omega_k\left (\hat{a}^+_{\mathbf{k},\sigma}\hat{a}_{\mathbf{k},\sigma}+\frac{1}{2}\right ),
\label{eq27}    
\end{equation}
so that the interaction Hamiltonian in the Coulomb gauge, when $\boldsymbol{\nabla}\cdot\mathbf{A}=0$ and hence $\hat{\mathbf{p}}_\alpha\cdot\mathbf{A}=\mathbf{A}\cdot\hat{\mathbf{p}}_\alpha$, is
\begin{equation}
    \hat{H}^{(A)}_{int}=-\sum\limits_\alpha\frac{q_\alpha}{m_\alpha c}\mathbf{A}\cdot\hat{\mathbf{p}}_\alpha+\sum\limits_\alpha\frac{q_\alpha^2}{2m_\alpha c^2}\mathbf{A}\cdot\mathbf{A}.
\label{eq28}    
\end{equation}
Another equally popular method, originally proposed by G\"{o}ppert-Mayer \cite{Goeppert-Mayer_1931}, is to introduce an interaction Hamiltonian that depends only on gauge-invariant electric and magnetic fields. In the electric dipole approximation $\mathbf{A}(\mathbf{r},t)\approx \mathbf{A}(\mathbf{0},t)$, and such a Hamiltonian can be obtained from (\ref{eq27}) by a unitary transformation \cite{Cohen-Tannoudji_1989}
\begin{equation}
\hat{H}^{(E)}= e^S\,\hat{H}^{(A)}\, e^{-S},\;\;S=-\frac{i}{\hbar c}\sum\limits_\alpha q_\alpha \mathbf{r}_\alpha\cdot\mathbf{A}(\mathbf{0},t)= -\frac{i}{\hbar c}\,\mathbf{d}\cdot\mathbf{A}(\mathbf{0},t),
\label{eq29}
\end{equation}
where $\mathbf{d}$ is the electric dipole moment of the system.

Obviously, the Coulomb potential energy $V_c$ in (\ref{eq27}) is invariant under such a transformation. To find the transformation properties of other terms in (\ref{eq27}), it is useful to
expand the vector potential in creation $\hat{a}^+_{\mathbf{k},\sigma}$ and annihilation $\hat{a}_{\mathbf{k},\sigma}$ operators
\begin{equation}
   \mathbf{A}=\sum\limits_{\mathbf{k},\sigma}\left(\frac{2\pi\hbar c^2}{L^3\omega_k}\right )^{1/2}\left( \hat{a}_{\mathbf{k},\sigma}\,e^{-i\omega_k t}\,e^{i\mathbf{k}\cdot\mathbf{r}}\,\boldsymbol{\epsilon}_{\mathbf{k},\sigma}+\hat{a}^+_{\mathbf{k},\sigma}\,e^{i\omega_k t}\,e^{-i\mathbf{k}\cdot\mathbf{r}}\,\boldsymbol{\epsilon}_{\mathbf{k},\sigma}\right),
\label{eq30} 
\end{equation}
where $\omega_k=kc$, $\boldsymbol{\epsilon}_{\mathbf{k},\sigma}$, $\sigma=1,2$ are two independent polarization vectors such that $\mathbf{k}\cdot\boldsymbol{\epsilon}_{\mathbf{k},\sigma}=0$, and $L^3$ is the quantization volume. As a result, we obtain \cite{Cohen-Tannoudji_1989}
\begin{equation}
    S=\sum\limits_{\mathbf{k},\sigma}\left (\lambda^*_{\mathbf{k},\sigma}\, \hat{a}_{\mathbf{k},\sigma}-\lambda_{\mathbf{k},\sigma}\, \hat{a}^+_{\mathbf{k},\sigma}\right),\;\;\lambda_{\mathbf{k},\sigma}=i\,\sqrt{\frac{2\pi}{\hbar L^3 \omega_k}}\,\,e^{i\omega_k t}\,\boldsymbol{\epsilon}_{\mathbf{k},\sigma}\cdot\mathbf{d}.
\label{eq31}    
\end{equation}
Using the following operator identity
\begin{equation}
    e^{\hat{A}} \hat{B} e^{-\hat{A}}=\hat{B}+[\hat{A},\hat{B}]+\frac{1}{2!}[\hat{A},[\hat{A},\hat{B}]]+
    \frac{1}{3!}[\hat{A}[\hat{A},[\hat{A},\hat{B}]]]+\cdots,
    \label{eq32}
\end{equation}
we get
\begin{equation}
 e^S\hat{\mathbf{p}}_\alpha e^{-S}= \hat{\mathbf{p}}_\alpha+[S,\hat{\mathbf{p}}_\alpha]=  \hat{\mathbf{p}}_\alpha+\frac{q_\alpha}{c}\mathbf{A},
 \label{eq33}
\end{equation}
where we used the canonical commutator $[(r_\alpha)_i,(p_\beta)_j]=i\hbar\,\delta_{\alpha\beta}\,\delta_{ij}$ and the fact that $[S,\hat{\mathbf{p}}_\alpha]=\frac{q_\alpha}{c}{\bf{A}}({\bf{0}},t)$ is a c-number (with respect to $\hat{\mathbf{p}}_\alpha$), so that all other commutators vanish. Another canonical commutator $[\hat{a}_{\mathbf{k},\sigma}(t),\hat{a}^+_{\mathbf{k}^\prime,\sigma^\prime}(t)]=\delta_{\mathbf{k}\mathbf{k}^\prime}\delta_{\sigma\sigma^\prime}$ in conjugation with (\ref{eq33}) gives
\begin{equation}
e^S\,\hat{a}_{\mathbf{k},\sigma}\,e^{-S}=\hat{a}_{\mathbf{k},\sigma}+\lambda_{\mathbf{k},\sigma}\,,\;\;
e^S\,\hat{a}^+_{\mathbf{k},\sigma}\,e^{-S}=\hat{a}^+_{\mathbf{k},\sigma}+\lambda^*_{\mathbf{k},\sigma}\,.
\label{eq34}    
\end{equation}
Using these last two results, we obtain for the new Hamiltonian
\begin{eqnarray} &&
\hat{H}^{(E)}= \sum\limits_\alpha\frac{1}{2m_\alpha}\hat{\mathbf{p}}_\alpha^2+V_c+\sum\limits_{\mathbf{k},\sigma}\hbar\omega_k\left (\hat{a}^+_{\mathbf{k},\sigma}\hat{a}_{\mathbf{k},\sigma}+\frac{1}{2}\right )+ \nonumber \\ &&
\sum\limits_{\mathbf{k},\sigma}\hbar\omega_k\left (\lambda^*_{\mathbf{k},\sigma}\hat{a}_{\mathbf{k},\sigma}+\lambda_{\mathbf{k},\sigma}\hat{a}^+_{\mathbf{k},\sigma}\right)+
\sum\limits_{\mathbf{k},\sigma}\hbar\omega_k\,\lambda^*_{\mathbf{k},\sigma}\lambda_{\mathbf{k},\sigma}\,.
\label{eq35}    
\end{eqnarray}
Since $\mathbf{E}=-\frac{1}{c}\frac{\partial \mathbf{A}}{\partial t}$, from (\ref{eq30}) and explicit expression for $\lambda_{\mathbf{k},\sigma}$ we easily get
\begin{equation}
\hat{H}^{(E)}_{int}=\sum\limits_{\mathbf{k},\sigma}\hbar\omega_k\left (\lambda^*_{\mathbf{k},\sigma}\hat{a}_{\mathbf{k},\sigma}+\lambda_{\mathbf{k},\sigma}\hat{a}^+_{\mathbf{k},\sigma}\right)=-\mathbf{d}\cdot\mathbf{E}(\boldsymbol{0},t),
\label{eq36}
\end{equation}
and 
\begin{equation}
\sum\limits_{\mathbf{k},\sigma}\hbar\omega_k\,\lambda^*_{\mathbf{k},\sigma}\lambda_{\mathbf{k},\sigma}=\sum\limits_{\mathbf{k},\sigma}\frac{2\pi}{L^3}\left(\boldsymbol{\epsilon}_{\mathbf{k},\sigma}\cdot\mathbf{d}\right)^2.
\label{eq37}
\end{equation}
The last expression depends only on the dipole moment $\mathbf{d}$ and represents the self-energy of the system in the dipole approximation. Formally, it diverges, but it should be remembered that the sum should be limited to the values of $\mathbf{k}$ for which the long-wave approximation is valid. Usually this self-interaction term is ignored. However, sometimes such self-energy terms are important for obtaining gauge-independent results, for example when calculating the Lamb shift \cite{Steck_2007}.

The Hamiltonians $\hat{H}^{(A)}$ and $\hat{H}^{(E)}$ are unitarily equivalent, and so they are expected to describe the same physics. It therefore came as a great surprise that Lamb, in his precise experiments on the Lamb-shift line shape, found that the experimental line shape agreed well with $\hat{H}^{(E)}$ but not with $\hat{H}^{(A)}$ \cite{Lamb_1952}. This apparent paradox has caused much discussion and confusion \cite{Cohen-Tannoudji_1989,Steck_2007,Power_1978,Schlicher_1984,Andrews_2018,Funai_2019,Stokes_2020}. The situation is eloquently described by Jaynes' comment ``How is it possible that a theory, for which formal
gauge invariance is proved easily once and for all, can lead to grossly non-invariant results as soon as we try to apply it to the simplest real problem?"  \cite{Kobe_1978}.

The problem can be explained as follows \cite{Lamb_1987}. Let us assume that a two-level atom with a resonant frequency $\omega_r=(E_e-E_g)/\hbar$ is excited by a plane electromagnetic wave with frequency $\omega$, so that $\mathbf{A}=\mathbf{A}_0e^{-i\omega t}$. Then $\mathbf{E}=-\frac{1}{c}\frac{\partial \mathbf{A}}{\partial t}=i\frac{\omega}{c}\bf{A}$. Therefore,
\begin{equation}
\langle e|-\frac{q}{mc}{\bf{A}}\cdot{\bf{\hat{p}}}\,|g\rangle =-\frac{q}{\hbar \omega}\langle e|\hat{H}_0\,{\bf{E}}\cdot{\bf{r}}-{\bf{E}}\cdot{\bf{r}}\,\hat{H}_0\,|g\rangle=\frac{\omega_r}{\omega}\langle e|-q {\bf{E}}\cdot{\bf{r}}\,|g\rangle\,,
\label{eq38}
\end{equation}
where we have used ${\bf{\hat{p}}}=\frac{im}{\hbar}[\hat{H}_0,{\bf{r}}]$, for $\hat{H}_0=\frac{{\bf{p}}^2}{2m}+V_c$, and the fact that in the Coulomb gauge $\bf{\hat{p}}\cdot{\bf{A}}={\bf{A}}\cdot\bf{\hat{p}}$ and hence $\bf{\hat{p}}\cdot{\bf{E}}={\bf{E}}\cdot\bf{\hat{p}}$. Paradoxically, for non-resonant (off-shell) transitions, the matrix elements in the $A$- and $E$-gauges give results that differ by a factor of $\omega_r/\omega$, which is small for narrow resonances, but was discovered in Lamb's precise experiments.

The standard resolution of this apparent paradox \cite{Bykov:1984,Lamb_1987} states that in order for gauge invariance to yield the same matrix elements, it is necessary to transform not only the interaction Hamiltonian, but also the initial and final states. If we transform only the interaction Hamiltonian, as in (\ref{eq38}), then of course we can get different results.

This explanation, although formally correct, misses the subtleties of the physical interpretation of states and observable operators in different gauges \cite{Steck_2007}. For example, suppose we want to calculate the matrix element of spontaneous light emission. Then the natural initial state is the photon-less state. However, (\ref{eq31}) points out that the operator $e^S$ transforms such a state into a coherent photon state. Why should we use such a complicated initial state in a different gauge \cite{Bykov:1984}?

Apparently, the problems that often accompany the use of the Coulomb gauge are rooted in the fact that charged particles are always accompanied by their own Coulomb fields. Thus, a gauge-invariant electron creation operator entails the simultaneous creation of an electron and the surrounding Coulomb field \cite{Dirac_1955}, and an incorrect treatment of the soft-photon Coulomb cloud around the electron can lead to infrared problems in quantum electrodynamics \cite{Kulish_1970}. 

It turned out that the $E$-gauge (and its multi-pole generalizations) is much less susceptible to such difficulties and allows one to perform high-order calculations with correct results more easily than the $A$-gauge, although, of course, with due care, correct results can be obtained in the $A$-gauge as well \cite{Power_1978,Gustin_2025}.

A generalization of the unitary transformation (\ref{eq29}) was given by Power and Zienau \cite{Power_1959}  and later by Woolley \cite{Wooley_1971}. It has the form
\begin{equation}
S=-\frac{i}{\hbar c}\int {\bf{P}}({\bf{r}})\cdot {\bf{A}}({\bf{r}})\,d{\bf{r}},\;\;  {\bf{P}}({\bf{r}})=\sum\limits_\alpha q_\alpha{\bf{r}}_\alpha\int\limits_0^1e^{-\lambda\,{\bf{r}}_\alpha\cdot{\boldsymbol{\nabla}}} \delta({\bf{r}})\,d\lambda.
\label{eq39}
\end{equation}
If only the first two terms are preserved in the multipole expansion
\begin{equation}
{\bf{P}}({\bf{r}})= \sum\limits_\alpha q_\alpha{\bf{r}}_\alpha\left [1-\frac{1}{2!}\left({\bf{r}}_\alpha\cdot\boldsymbol{\nabla}\right)+\frac{1}{3!}\left({\bf{r}}_\alpha\cdot\boldsymbol{\nabla}\right)^2-\frac{1}{4!}\left({\bf{r}}_\alpha\cdot\boldsymbol{\nabla}\right)^3+\cdots\right] \delta({\bf{r}}),
\label{eq40}
\end{equation}
we obtain $S=S^{(0)}+S^{(1)}$, where $S^{(0)}$ is given by (\ref{eq29}) or (\ref{eq31}), while
\begin{equation}
S^{(1)}=-\frac{i}{2\hbar c}\sum\limits_\alpha q_\alpha \left .\left({\bf{r}}_\alpha\cdot\boldsymbol{\nabla}\right) \left({\bf{r}}_\alpha\cdot{\bf{A}}\right)\,\right |_{{\bf{r}}={\bf{0}}}= \sum\limits_{\mathbf{k},\sigma}\left (\mu^*_{\mathbf{k},\sigma}\, \hat{a}_{\mathbf{k},\sigma}-\mu_{\mathbf{k},\sigma}\, \hat{a}^+_{\mathbf{k},\sigma}\right), 
\label{eq41}
\end{equation}
with
\begin{equation}
\mu_{\mathbf{k},\sigma}=\left(\frac{2\pi}{L^3\hbar\omega_k}\right)^{1/2}e^{i\omega_kt}\, {\bf{k}}\cdot{\bf{Q}}\cdot\boldsymbol{\epsilon}_{{\bf{k}},\sigma}\,,\;\;Q_{i,j}=\frac{1}{2}\sum\limits_\alpha q_\alpha(r_\alpha)_i(r_\alpha)_j,\;\;i,j=x,y,z,
\label{eq42}
\end{equation}
 where $Q_{i,j}$ is the electric quadrupole moment of the system.
 
 When calculating the transformed Hamiltonian $\hat{H}^{(E)}$, we must take into account that ${\bf{A}}({\bf{r}}_\alpha,t)$ in $\hat{H}^{(A)}$ must be expanded with the same accuracy as in the operator $S$: ${\bf{A}}({\bf{r}}_\alpha,t)\approx {\bf{A}}({\bf{0}}_\alpha,t)+\left ({\bf{r}}_\alpha\cdot \boldsymbol{\nabla}\right )\left . {\bf{A}}\right |_{{\bf{r}}=0}$. Relevant commutators are
\begin{equation}
[\nabla_iA_j|_{{\bf{r}}=0}, \nabla_iA_j|_{{\bf{r}}=0}]=0,\;\;[S^{(1)},{\hat{\bf{p}}_\alpha}]=\frac{q_\alpha}{2c}\left .\left [{\boldsymbol{\nabla}}\left({\bf{r}}_\alpha\cdot{\bf{A}}\right)+\left({\bf{r}}_\alpha\cdot\boldsymbol{\nabla}\right){\bf{A}}\right ]\right |_{{\bf{r}=0}},
\label{eq43}
\end{equation}
and
\begin{equation}
e^S\,\hat{a}_{\mathbf{k},\sigma}\,e^{-S}=\hat{a}_{\mathbf{k},\sigma}+\lambda_{\mathbf{k},\sigma}+\mu_{\mathbf{k},\sigma}\,,\;\;
e^S\,\hat{a}^+_{\mathbf{k},\sigma}\,e^{-S}=\hat{a}^+_{\mathbf{k},\sigma}+\lambda^*_{\mathbf{k},\sigma}+\mu^*_{\mathbf{k},\sigma}\,.
\label{eq44}    
\end{equation}
As a result, since ${\boldsymbol{\nabla}}\left({\bf{r}}_\alpha\cdot{\bf{A}}\right)-\left({\bf{r}}_\alpha\cdot\boldsymbol{\nabla}\right){\bf{A}}={\bf{r}}_\alpha\times\left(\boldsymbol{\nabla}\times{\bf{A}}\right)={\bf{r}}_\alpha\times{\bf{H}}$, we get
\begin{eqnarray} &&
\hat{H}^{(E)}=\sum\limits_\alpha\frac{1}{2m_\alpha}\left(\hat{{\bf{p}}}_\alpha+\frac{q_\alpha}{2c}\, {\bf{r}}_\alpha\times{\bf{H}}(\boldsymbol{0},t)\right )^2+V_c+ \nonumber \\ &&
\sum\limits_{{\bf{k}},\sigma}\hbar\omega_k\left((\hat{a}^+_{\mathbf{k},\sigma}+\lambda^*_{\mathbf{k},\sigma}+\mu^*_{\mathbf{k},\sigma})(\hat{a}_{\mathbf{k},\sigma}+\lambda_{\mathbf{k},\sigma}+\mu_{\mathbf{k},\sigma})+\frac{1}{2}\right).
\label{eq45}
\end{eqnarray}
Like to (\ref{eq36}), we obtain
\begin{equation}
\sum\limits_{\mathbf{k},\sigma}\hbar\omega_k\left (\mu^*_{\mathbf{k},\sigma}\hat{a}_{\mathbf{k},\sigma}+\mu_{\mathbf{k},\sigma}\hat{a}^+_{\mathbf{k},\sigma}\right)=-\boldsymbol{\nabla}\cdot{\bf{Q}}\cdot{\mathbf{E}}\,|_{\boldsymbol{r}=0}\equiv-\sum\limits_{ij}Q_{ij}\nabla_i E_j\,|_{\boldsymbol{r}=0}\,,
\label{eq46}
\end{equation}
where $Q_{ij}$ is the electric quadrupole moment of the system introduced in (\ref{eq42}). On the other hand
\begin{equation}
\sum\limits_\alpha  \frac{q_\alpha}{4m_\alpha c}\left[ \hat{{\bf{p}}}_\alpha\cdot ({\bf{r}}_\alpha\times{\bf{H}})+({\bf{r}}_\alpha\times{\bf{H}})\cdot \hat{{\bf{p}}}_\alpha  \right ]=- \sum\limits_\alpha  \frac{q_\alpha}{2m_\alpha c}({\bf{r}}_\alpha\times\hat{\bf{p}}_\alpha)\cdot{\bf{H}}=-\boldsymbol{\mu}\cdot{\bf{H}}\,, 
\label{eq47}
\end{equation}
where $\boldsymbol{\mu}$ is the magnetic dipole moment of the system. Note that ${\bf{r}}_\alpha\times\hat{\bf{p}}_\alpha=-\hat{\bf{p}}_\alpha\times\bf{r}_\alpha$, even though the operators involved do not commute. 

As a result, the Hamiltonian (\ref{eq45}) takes the form
\begin{eqnarray} &&  \hat{H}^{(E)}=\sum\limits_\alpha\frac{\hat{{\bf{p}}}_\alpha^2}{2m_\alpha}+V_c+\sum\limits_{{\bf{k}},\sigma}\hbar\omega_k\left(\hat{a}^+_{\mathbf{k},\sigma}\hat{a}_{\mathbf{k},\sigma}+\frac{1}{2}\right)+\hat{H}_{int}^{(E)}+\nonumber \\ && \sum\limits_\alpha\frac{q_\alpha^2}{8m_\alpha c}\left[{\bf{r}}_\alpha\times {\bf{H}}(\boldsymbol{0},t)\right ]^2+\sum\limits_{{\bf{k}},\sigma}\hbar\omega_k\left(\lambda^*_{\mathbf{k},\sigma}+\mu^*_{\mathbf{k},\sigma}\right)\left (\lambda_{\mathbf{k},\sigma}+\mu_{\mathbf{k},\sigma}\right)
\label{eq48}
\end{eqnarray}
with the multipole interaction Hamiltonian
\begin{equation}
\hat{H}_{int}^{(E)}=-\mathbf{d}\cdot\mathbf{E}(\boldsymbol{0},t)-\boldsymbol{\mu}\cdot{\bf{H}}(\boldsymbol{0},t)-\left(\boldsymbol{\nabla}\cdot{\bf{Q}}\cdot{\mathbf{E}}\right)(\boldsymbol{0},t)\,.
\label{eq49}
\end{equation}
For particles with spin we must of course add the corresponding contribution with the proper gyromagnetic factor to the magnetic moment $\mu$.

The last term in (\ref{eq48}) describes the field-independent self-interaction of electric multipoles and is a generalization of (\ref{eq37}). This term is important in calculations of the Lamb shift or van der Waals forces, but in most other cases it can be ignored \cite{Loudon_1983}.

One of the delicate aspects of the unitary transformation under consideration is the need to recognize that seemingly identical operators, in particular the canonical momenta, actually have different physical meanings in the original and derived Hamiltonians. We will not consider this interesting aspect further, suggesting that the interested reader can consult the literature \cite{Loudon_1983,Cohen_2019,Grynberg_2010}.

\section{Dicke superradiance: quantum considerations}
 For a two-level atom with excited $|e\rangle$ and ground $|g\rangle$ states, the interaction Hamiltonian (\ref{eq49}) in conjugation with Fermi's golden rule leads to the following emission rate of the electric dipole transition \cite{Harris_1972}:
 \begin{equation}
 \gamma=\frac{4\omega^3}{3\hbar c^3}\,\left |\langle g|\,{\bf{d}}\,|e\rangle \right|^2,
 \label{eq50}
 \end{equation}
where $\omega=(E_e-E_g)/\hbar$. A similar result can be obtained for the magnetic dipole transition. Indeed, for $\hat{H}_{int}=-\boldsymbol{\mu}\cdot{\bf{H}}(\boldsymbol{0},t)$, ${\bf{H}}=\boldsymbol{\nabla}\times{\bf{A}}$, and $|i\rangle=|e\rangle|0\rangle$, $|f\rangle=|g\rangle|{\bf{k}},\sigma\rangle$, Fermi's golden rule
\begin{equation}
 \gamma=\frac{2\pi}{\hbar}\sum\limits_{f-states} \left |\langle f|\hat{H}_{int}|i\rangle \right |^2 
 \delta(E_f-E_i)
 \label{eq51}
\end{equation}
will give
\begin{equation}
\gamma=\frac{(2\pi)^2c^2}{L^3}\sum\limits_{{\bf{k}},\sigma}\frac{1}{\omega_k}\left|\,\left({\bf{k}}\times{\bf{\epsilon}}_{{\bf{k}},\sigma}\right)\cdot\langle g|\,\boldsymbol{\mu}\,|e\rangle\,\right |^2 \delta(\hbar\omega-\hbar\omega_k)\,.
\label{eq52}
\end{equation}
Summation over ${\bf{k}}$ in the continuum limit $L\to \infty$ can be replaced by integration \cite{Harris_1972}:
\begin{equation}
\frac{1}{L^3}\sum\limits_{\bf{k}} \to \int\frac{d{\bf{k}}}{(2\pi)^3}=\int\frac{k^2dk\,d\Omega_{\bf{k}}}{(2\pi)^3},
\label{eq53}
\end{equation}
while the sum over photon polarizations $\sigma$ can be performed using
\begin{equation}
\sum\limits_{\sigma}\left(\epsilon_{{\bf{k},\sigma}}\right )_i\, \left(\epsilon_{{\bf{k},\sigma}}\right )_j=\delta_{ij}-\frac{k_i\,k_j}{k^2}.
\label{eq54}
\end{equation}
As a result we get
\begin{equation}
\gamma=\left .\frac{\omega}{2\pi c\hbar}\int d\Omega_{\bf{k}}\,\left[\, {\bf{k}}\times\langle g|\,\boldsymbol{\mu}\,|e\rangle\,\right ]^2\right|_{k=\frac{\omega}{c}}=\frac{4\omega^3}{3\hbar c^3}\,
\left |\,\langle g|\,\boldsymbol{\mu}\,|e\rangle\,\right |^2
\label{eq55}
\end{equation}
in complete analogy with (\ref{eq50}).

An important point that follows from the condition of validity of the dipole approximation is that (\ref{eq50}) and (\ref{eq55}) are valid for any quantum system whose size is much smaller than the wavelength of the radiation. Following Dicke \cite{Dicke_1954}, we apply these formulas to a system of $N$ two-level atoms, which form one quantum system due to the close arrangement of the atoms. The fact that the radiation comes from a specific quantum state of the system, and not from individual atoms, completely changes the nature of the radiation. In particular, a quantum state symmetric with respect to the atoms leads to an extremely strong and short burst of radiation of the type (\ref{eq17}), called Dicke superradiance.  We will briefly outline only the elementary theory of this phenomenon \cite{Ermolaev_1996}. Many other interesting and important details about Dicke superradiance and related phenomena at a deeper and more realistic level can be found in the extensive literature, from which we cite only a few \cite{Gross_1982,Zheleznyakov_1989,Menshikov_1999,Ermolaev_1996,Eberly_1987}.

The quantum state of a two-level atom $|\psi\rangle=\alpha |e\rangle+\beta |g\rangle$ can be represented by a column matrix with two rows: $$\psi=\left(\begin{array}{c} \alpha \\ \beta \end{array}\right),$$
while the quantum mechanical operators will be represented through $2\times 2$ square matrices. In particular, if the origin of the energy scale is chosen at the midpoint of the energy interval between the excited and ground states, then the free atomic Hamiltonian will be
\begin{equation}
\hat{H}_0=\left (\begin{array}{cc} \frac{1}{2}\hbar\omega & 0 \\ 0 &  -\frac{1}{2}\hbar\omega\end{array}\right)=\hbar\omega\hat{R}_3,\;\; \hat{R}_3=\frac{1}{2}\sigma_3.  
\end{equation}
The pseudo-spin operators associated with the Pauli matrices
\begin{equation}
\hat{R}_1=\frac{1}{2}\sigma_1,\;\; \hat{R}_2=\frac{1}{2}\sigma_2, \;\;\hat{R}_3=\frac{1}{2}\sigma_3
\label{eq57}
\end{equation}
satisfy the usual commutation relations of angular momentum $[\hat{R}_m,\,\hat{R}_n]=i\hat{R}_l$, where $(m,n,l)$ is any cyclic permutation of the numbers $1,2,3$. In terms of these operators, the Hamiltonian of the interaction of an atom with light has the form (assuming $\Delta m=0$ transition, so that ${\bf{d}}_{eg}={\bf{d}}_{ge}\equiv{\bf{d}}$ and similarly for the magnetic dipole transition)
\begin{eqnarray}  &&
\hat{H}_{int}=-\mathbf{E}(\boldsymbol{0},t)\cdot\left (\begin{array}{cc} 0 & {\bf{d}} \\ {\bf{d}} & 0\end{array}\right)=-2\mathbf{d}\cdot\mathbf{E}(\boldsymbol{0},t)\hat{R}_1,\nonumber \\ &&
\hat{H}_{int}=-\mathbf{H}(\boldsymbol{0},t)\cdot\left (\begin{array}{cc} 0 & {\boldsymbol{\mu}} \\ {\boldsymbol{\mu}} & 0\end{array}\right)=-2\boldsymbol{\mu}\cdot\mathbf{H}(\boldsymbol{0},t)\hat{R}_1,
\label{eq58}
\end{eqnarray}
for electric dipole and magnetic dipole interactions, respectively.

For an ensemble of $N$-identical atoms, we introduce individual pseudo-spin operators $\hat{R}^{(\alpha)}$ with $[\hat{R}^{(\alpha)}_m,\,\hat{R}^{(\beta)}_n]=i\delta_{\alpha\beta}\,\hat{R}^{(\alpha)}_l$, in terms of which the Hamiltonian of the atoms+electromagnetic field system takes the form
\begin{eqnarray} &&
\hat{H}=\hbar\omega\sum\limits_\alpha\hat{R}^{(\alpha)}_3-2g \sum\limits_\alpha\hat{R}^{(\alpha)}_1+\sum\limits_{{\bf{k}},\sigma}\hbar\omega_k\left(\hat{a}^+_{\mathbf{k},\sigma}\hat{a}_{\mathbf{k},\sigma}+\frac{1}{2}\right)= \nonumber \\ && \hbar\omega\,\hat{R}_3-2g\,\hat{R}_1+\sum\limits_{{\bf{k}},\sigma}\hbar\omega_k\left(\hat{a}^+_{\mathbf{k},\sigma}\hat{a}_{\mathbf{k},\sigma}+\frac{1}{2}\right),
\label{eq59}
\end{eqnarray}
where $g=\mathbf{d}\cdot\mathbf{E}(\boldsymbol{0},t)$ or $g=\boldsymbol{\mu}\cdot\mathbf{H}(\boldsymbol{0},t)$, and we introduced the total quasi-spin operators of the $N$-atom system
\begin{equation}
\hat{R}_i=\sum\limits_{\alpha=1}^N  \hat{R}^{(\alpha)}_i,\;\;i=1,2,3;\;\;\;[\hat{R}_m,\,\hat{R}_n]=i\sum\limits_{l=1}^3\epsilon_{mnl}\,\hat{R}_l\, .
\label{eq60}
\end{equation}
From the angular momentum commutation relations it follows that $\hat{R}^2=\hat{R}_1^2+\hat{R}_2^2+\hat{R}_3^2$ commutes with all pseudo-spin operators and, consequently, with $\hat{H}$. Indeed,
$$[\hat{R}^2,\hat{R}_m]=\sum\limits_{n=1}^3\left(\hat{R}_n[\hat{R}_n,\hat{R}_m]+[\hat{R}_n,\hat{R}_m]\hat{R}_n\right)=i\sum\limits_{n,l=1}^3\epsilon_{nml}\left(\hat{R}_n\hat{R}_l+\hat{R}_l\hat{R}_n\right)=0,$$
since the first factor is antisymmetric with respect to $n,l$, and the second is symmetric.

Therefore, the quantum state of the ensemble $|J,M\rangle$ is characterized by the total pseudo-spin $J$ (the eigenvalue of $\hat{R}^2$ in this state is $J(J+1)$), and by $M$, the eigenvalue of $\hat{R}_3$. It is clear from its definition, that $M=\frac{1}{2}\left(N_e-N_g\right)$, where $N_e$ is the number of atoms in the excited state, and $N_g$ is the number of atoms in the ground state. If initially all atoms are in the excited state, $M=\frac{1}{2}N,\frac{1}{2}N-1,\frac{1}{2}N-2,\ldots,-\frac{1}{2}N$, and $J=\frac{1}{2}N$.

Since $2\hat{R}_1=\hat{R}_++\hat{R}_-$, where $\hat{R}_+=\hat{R}_1+i\hat{R}_2$ and $\hat{R}_-=\hat{R}_1-i\hat{R}_2$ are raising and lowering operators, and, as follows from the commutation relations of angular momentum, for normalized states
\begin{eqnarray} &&
\hat{R}_+\,|J,M\rangle=\sqrt{(J-M)(J+M+1)}\,|J,M+1\rangle, \nonumber \\ &&
\hat{R}_-\,|J,M\rangle=\sqrt{(J+M)(J-M+1)}\,|J,M-1\rangle,
\label{eq61}
\end{eqnarray}
radiative transition occurs only between neighboring $|J,M\rangle$ states. At that, for transition $|J,M\rangle\to|J,M-1\rangle$ the matrix element of the total dipole moment is 
\begin{equation}
\mu_{M,M-1}=g\,\langle J,M-1|\left (\hat{R}_++\hat{R}_-\right)|J,M\rangle=g\,\sqrt{(J+M)(J-M+1)}.
\label{eq62}
\end{equation}
Therefore, assuming that the decay cascade starts from the state $|\frac{N}{2},\frac{N}{2}\rangle$, according to (\ref{eq50}) (or according to (\ref{eq55}) if the transition is of the magnetic dipole type), the corresponding decay rate is
\begin{equation}
\gamma_{M,M-1}=\gamma \left (\frac{N}{2}+M\right)\left(\frac{N}{2}-M+1\right ).
\label{eq63}
\end{equation}
This expression reaches its maximum in the middle of the decay cascade, when $M=1\,\mathrm{or}\,0$, if $N$ is even, or when $M=\frac{1}{2}$, if $N$ is odd. If $N\gg 1$, this maximum rate is proportional to $N^2$, indicating the onset of coherent emission. At the beginning of the decay cascade, when $M\approx\frac{N}{2}$, the rate is proportional to $N$, i.e. the atoms begin to decay incoherently, and coherence only emerges later due to their common coupling to a resonant mode of the electromagnetic field.

Let us assume that the system is initially completely inverted (excited) and that at time $t(N_g)$ the $N_g$ atoms have already returned to the ground state (that is, by this moment $N_g$ photons have been emitted). We can approximate $t(N_g)$ by the sum of the inverse rates $\gamma^{-1}_{M,M-1}$, that is, the times of individual cascade transitions \cite{Ermolaev_1996}. It follows from $N=N_e+N_g$ and $2M=N_e-N_g$ that $M(N_g)=\frac{N}{2}-N_g$. Therefore, in the last cascade transition $M\to M-1$ we must have $M=\frac{N}{2}-N_g+1$ to end with $M-1=\frac{N}{2}-N_g=M(N_g)$ and we get the formula
\begin{equation}
    t(N_g)=\sum\limits_{M=\frac{N}{2}-N_g+1}^{\frac{N}{2}}\gamma^{-1}_{M,M-1}=\sum\limits_{M=\frac{N}{2}-N_g+1}^{\frac{N}{2}}\frac{\tau_0}{\left (\frac{N}{2}+M\right)\left(\frac{N}{2}-M+1\right )},
\label{eq64}
\end{equation}
where $\tau_0=\gamma^{-1}$ is the decay constant of a single atom. The sum can be calculated approximately using the Euler-Maclaurin formula \cite{Kac_2002}
\begin{eqnarray} &&\hspace*{-5mm}
\sum\limits_{M=M_1}^{M_2-1}f(M)-\int\limits_{M_!}^{M_2}f(x)dx=-\frac{1}{2}\left [f(M_2)-f(M_1)\right ]+   \nonumber\\&& \hspace*{-5mm}
\sum\limits_{k=1}^m\frac{B_{2k}}{(2k)!}\,\left [f^{(2k-1)}(M_2)-f^{(2k-1)}(M_1)\right ]-\int\limits_{M_1}^{M_2}\frac{b_{2m+1}(\{1-x\})}{(2m+1)!}\,f^{(2m+1)}(x)\,dx,
\label{eq65}
\end{eqnarray}
where $B_{2k}$ are the Bernoulli numbers, $b_{2m+1}(x)$ are the Bernoulli polynomials, $\{x\}$ denotes the fractional part of a real number $x$, and $m$ is some integer. The last, so-called remainder term, can be made small by choosing an appropriate $m$. Namely, the following estimate holds \cite{Kac_2002}
\begin{equation}
 \left | \int\limits_{M_1}^{M_2}\frac{b_{2m+1}(\{1-x\})}{(2m+1)!}\,f^{(2m+1)}(x)\,dx \right |<\frac{4e^{2\pi}}{(2\pi)^{2m+1}}  \int\limits_{M_1}^{M_2}\left|f^{(2m+1)}(x)\right|\,dx.
 \label{eq66}
\end{equation}

To use the Euler-Maclaurin formula in our case, we rewrite
\begin{equation}
t=\frac{\tau_0}{N}+\sum\limits_{M=\frac{N}{2}-N_g+1}^{\frac{N}{2}-1}f(M),\;\;f(M)=\frac{\tau_0}{N+1}\left[\frac{1}{\frac{N}{2}+M}+\frac{1}{\frac{N}{2}-M+1}\right].
\label{eq67}
\end{equation}
Then after trivial integration we get
\begin{equation}
 t=\frac{\tau_0}{N}+\frac{\tau_0}{N+1}\ln{\frac{NN_g}{N-N_g+1}}+\frac{\tau_0r}{N+1},
 \label{eq68}
\end{equation}
where $r\sim O(1)$ arises from the right-hand side of (\ref{eq65}). Assuming $N\gg 1$, we rewrite (\ref{eq68}) as follows
\begin{equation}
 \frac{N}{\tau_0}t-\ln{N}-1-r=\ln{\frac{N_g}{N-N_g}}.
 \label{eq69}
\end{equation}
To logarithmic accuracy, $1+r$ can be neglected compared to $\ln{N}$, and after solving (\ref{eq69}) for $N_g$ we obtain
\begin{equation}
N_g=N\,\frac{e^{\frac{N}{\tau_0}(t-t_0)}}{e^{\frac{N}{\tau_0}(t-t_0)}+1},\;\;t_0=\frac{\ln{N}}{N}\,\tau_0.
\label{eq70}
\end{equation}
Then
\begin{equation}
M=\frac{N}{2}-N_g=-\frac{N}{2}\,\tanh{\left [\frac{N}{2\tau_0}(t-t_0)\right]}.
\label{71}
\end{equation}
This is effectively the same relation as (\ref{eq16}) (with $\lambda=1$), since the dimensionless energy $W$ (in units of $\hbar\omega$) stored in the system is $W=\frac{\hbar\omega}{2}(N_e-N_g)/(\hbar\omega)=M$.

\section{Dicke superradiance from large samples} 
Dicke model in the previous section assumes ideal conditions: 
\begin{itemize}
    \item all atoms are located in a volume small compared to $\lambda^3$;
    \item  however, neighboring atoms are far enough away that there is no need to worry about atom-atom interactions;
    \item there is complete permutation symmetry in the interaction of atoms with their common electromagnetic field;
    \item emitted photons immediately flow out of the sample, so any back-reaction on their part and propagation effects are ignored;
    \item atomic levels of individual atoms involved are non-degenerate;
    \item collisions between atoms do not affect their internal states, and collisional broadening of atomic levels is not taken into account;
    \item it is assumed that the radiation field is uniform throughout the sample.
\end{itemize}    
In practice, and especially under natural circumstances, all these conditions are difficult to fulfill.
However, Dicke superradiance is possible even for large samples \cite{Dicke_1954}.

First, a very simple interpolation is possible between the quantum incoherent exponential decay law and the classical coherent radiation pulse (\ref{eq17}) \cite{Eberly_1972}. When initially fully inverted $N$ two-level atoms decay incoherently, the exponential decay law is $N_e=Ne^{-t/\tau_0}$, or
\begin{equation}
W=M=\frac{1}{2}(N_e-N_g)=N_e-\frac{N}{2}=Ne^{-t/\tau_0}-\frac{N}{2}.
\label{eq72}
\end{equation}
Therefore,
\begin{equation}
-\frac{dW}{dt}=\frac{1}{\tau_0} Ne^{-t/\tau_0}= \frac{1}{\tau_0}\left (\frac{N}{2}+W\right ).
\label{eq73}
\end{equation}
This should be compared with the classical law of coherent decay (\ref{eq14}). The interpolation formula between these two extremes is \cite{Eberly_1972}
\begin{equation}
-\frac{dW}{dt}=\frac{\lambda}{\tau_0}\left (\frac{N}{2}+W\right ) \left (\frac{N}{2}-W+\frac{1}{\lambda} \right ).
\label{eq74}
\end{equation}
Initially, when $W\approx N/2$, $N/2-W\ll 1/\lambda$ and (\ref{eq74}) transforms into the exponential decay law (\ref{eq73}). While later, when at $W\approx 0$ the macroscopic dipole moment accumulates, $N/2-W\gg 1/\lambda$ and (\ref{eq74}) transforms into the classical radiation law (\ref{eq14}).

If $t_0$ is the time when $-dW/dt$ in (\ref{eq74}) reaches its maximum value at $W_0=\frac{1}{2\lambda}$, then the corresponding solution of the differential equation (\ref{eq74}) is
\begin{equation}
W=\frac{N}{2}\left [-\left(1+\frac{1}{N\lambda}\right)\tanh{\left(\frac{N\lambda+1}{2\tau_0}\,(t-t_0)\right )}+\frac{1}{N\lambda}\right],
\label{eq75}
\end{equation}
that corresponds to the radiation intensity \cite{Eberly_1972}
\begin{equation}
I(t)=-\hbar\omega\,\frac{dW}{dt}=\hbar\omega\,\frac{(N\lambda+1)^2}{4\lambda\tau_0}\,\sech^2{\left(\frac{N\lambda+1}{2\tau_0}(t-t_0)\right)}.
\label{eq76}
\end{equation}
A more rigorous quantum mechanical justification of the interpolation formula (\ref{eq74}) and the results following from it can be found in \cite{Eberly_1971}.

From (\ref{eq75}) and (\ref{eq76}) it is clear that superradiance is characterized by two time scales
\begin{equation}
T_{SR}=\frac{\tau_0}{N\lambda+1},\;\;t_0=T_{SR}\,\ln{(N\lambda)}.
\label{eq77}
\end{equation}
The expression for $t_0$ follows from the initial condition $W(t=0)=N/2$ and represents the delay time, the time during which macroscopic coherence is established and the radiation intensity increases to its superradiant peak of short duration $T_{SR}$.

The product $N\lambda$ can be considered as a parameter characterizing the superradiance of a large system, which plays the same role as the total number of atoms $N$ in the dynamics of a small system. The shape factor $\lambda$ is a complex function of the size and shape of the volume containing the two-level atoms. However, in two extreme cases for samples in the form of a needle-like long cylinder of length $L$ and in the form of a flat disk of radius $R$, the following estimates are valid \cite{Eberly_1971}
\begin{equation}
\lambda_C\approx \frac{3\pi}{4}\,\frac{c}{\omega L},\;\;\;\lambda_D\approx3\left(\frac{c}{\omega R}\right )^2.   
\label{eq78}    
\end{equation}
The dynamics of Dicke superradiance of a large sample can be conveniently described by the semiclassical Maxwell-Bloch equations, which can be obtained as follows \cite{Ermolaev_1996}. As a first step, we introduce the density matrix $\hat\rho=\psi\psi^+$ for a two-level atom with wave function $\psi$:
\begin{equation}
 \psi=\left(\begin{array}{c} \alpha \\ \beta \end{array}\right),\;\;\hat\rho= \left(\begin{array}{cc} \rho_{ee} &\rho_{eg} \\ \rho_{ge} & \rho_{gg} \end{array}\right)=\left(\begin{array}{cc} \alpha\alpha^* &\alpha\beta^* \\ \beta\alpha^* & \beta\beta^* \end{array}\right) .
 \label{eq79}
\end{equation}
For an ensemble of $N$ two-level atoms, the density matrix has the form 
\begin{equation}
\hat\rho(\boldsymbol{r,t)}=\frac{1}{N}\sum\limits_{\alpha=1}^N\delta(\boldsymbol{r}-\boldsymbol{r}_\alpha)\hat\rho^{(\alpha)}(t),
\label{eq80} 
\end{equation}
where $\hat\rho^{(\alpha)}$ are the density matrices of individual atoms, averaged over volumes large compared to the size of the atom and small compared to the emitted wavelength \cite{Gross_1982}. According to (\ref{eq59}), the Schr\"{o}dinger equation for the wave function $\psi$ is
\begin{equation}
 i\hbar\, \frac{\partial}{\partial t}\, \left(\begin{array}{c} \alpha \\ \beta \end{array}\right)= \left(\begin{array}{cc} \frac{1}{2}\hbar\omega &-g \\ -g & -\frac{1}{2}\hbar\omega \end{array}\right) \left(\begin{array}{c} \alpha \\ \beta \end{array}\right).
 \label{eq81}
\end{equation}
From (\ref{eq79}) and (\ref{eq81}) it follows that the components of the density matrix satisfy a simple precession-type equation \cite{Feynman_1957,Eberly_1987}
\begin{equation}
\frac{d\boldsymbol{r}}{dt}=\boldsymbol{\Omega}\times \boldsymbol{r},\;\; \boldsymbol{r}= \Big(\rho_{eg}+\rho_{ge},\,i(\rho_{eg}-\rho_{ge}),\,\rho_{ee}-\rho_{gg}\Big),\;\; \boldsymbol{\Omega}=\left(-\frac{2g}{\hbar},0,\omega\right ).
\label{eq82}
\end{equation}
Since $\rho_{ee}+\rho_{gg}=1$, we get from (\ref{eq82}) 
\begin{eqnarray}  &&
\frac{\partial \rho_{ee}}{\partial t}=i\frac{g}{\hbar}(\rho_{ge}-\rho_{eg}),\;\; \frac{\partial \rho_{gg}}{\partial t}=-i\frac{g}{\hbar}(\rho_{ge}-\rho_{eg}),\nonumber\\ && \frac{\partial \rho_{eg}}{\partial t}=-i\omega \,\rho_{eg}-i\frac{g}{\hbar}(\rho_{ee}-\rho_{gg}),\;\;\rho_{ge}=\rho_{eg}^*.
\label{eq83}
\end{eqnarray}
These equations can be also obtained from the von Neumann equation for the density matrix \cite{Ermolaev_1996}. 

Since $g=\mathbf{d}\cdot\mathbf{E}(\boldsymbol{r},t)$ or $g=\boldsymbol{\mu}\cdot\mathbf{H}(\boldsymbol{r},t)$, to close the system (\ref{eq83}) we need the equation of evolution of the electromagnetic field, which can be obtained from Maxwell's macroscopic equations:
\begin{eqnarray} &&
\boldsymbol{\nabla}\cdot{\bf{D}}=4\pi\rho_f,\;\; \boldsymbol{\nabla}\cdot{\bf{B}}=0,\;\;{\bf{D}}={\bf{E}}+4\pi\,{\bf{P}},\;\;{\bf{B}}={\bf{H}}+4\pi\,{\bf{M}}, \nonumber \\ &&
\boldsymbol{\nabla}\times{\bf{H}}=\frac{4\pi}{c}\,{\bf{J}}_f+\frac{1}{c}\frac{\partial {\bf{D}}}{\partial t},\;\; \;\; \boldsymbol{\nabla}\times{\bf{E}}=-\frac{1}{c}\frac{\partial {\bf{B}}}{\partial t}.
\label{eq84}
\end{eqnarray}
Assuming no free charges ($\rho_f=0$, ${\bf{J}}_f={\bf{0}}$), and ${\bf{M}}=0$, $\boldsymbol{\nabla}\cdot{\bf{P}}=0$ in case of electric dipole transitions, so that $\boldsymbol{\nabla}\times(\boldsymbol{\nabla}\times{\bf{E}})=\boldsymbol{\nabla}(\boldsymbol{\nabla}\cdot{\bf{E}})-\boldsymbol{\nabla}^2{\bf{E}}$, we get from (\ref{eq84})
\begin{equation}
\boldsymbol{\nabla}^2{\bf{E}} \,-\frac{1}{c^2}\frac{\partial^2{\bf{E}}}{\partial t^2}=\frac{4\pi}{c^2}\frac{\partial^2{\bf{P}}}{\partial t^2}.
\label{eq85}
\end{equation}
Analogously, for magnetic dipole transitions, assuming ${\bf{P}}=0$, $\boldsymbol{\nabla}\cdot{\bf{M}}=0$, we obtain
\begin{equation}
\boldsymbol{\nabla}^2{\bf{H}} \,-\frac{1}{c^2}\frac{\partial^2{\bf{H}}}{\partial t^2}=\frac{4\pi}{c^2}\frac{\partial^2{\bf{M}}}{\partial t^2}.
\label{eq86}
\end{equation}
The polarization density ${\bf{P}}$ can be viewed as the quantum mechanical expectation value of the electric dipole moment per unit volume. Similarly, the magnetization density ${\bf{M}}$ can be treated as the quantum mechanical expectation value of the magnetic dipole moment per unit volume. Therefore \cite{Ermolaev_1996},
\begin{equation}
{\bf{P}}=n{\bf{d}}\,\mathrm{Tr}( 2\hat{\rho}\hat{R}_1)= n{\bf{d}}\,(\rho_{eg}+\rho_{ge}),\;\;
{\bf{M}}=n{\boldsymbol{\mu}}\,\mathrm{Tr}( 2\hat{\rho}\hat{R}_1)= n{\boldsymbol{\mu}}\,(\rho_{eg}+\rho_{ge}),
\label{eq87}
\end{equation}
where $n$ is the number of atoms per unit volume.

For a sample in the form of a needle-like long cylinder along the $z$ axis, the solution of (\ref{eq83}) and (\ref{eq85}) can be sought in the form of linearly polarized (with real polarization vector $\boldsymbol{\epsilon}_k$) plane waves propagating in the positive direction of the $z$ axis \cite{Ermolaev_1996}:
\begin{eqnarray} &&
\rho_{eg}=\frac{1}{2}R(z,t)e^{-i(\omega t-kz)}, \nonumber \\ &&{\bf{E}}({\bf{r}},t)=\frac{1}{2}\boldsymbol{\epsilon}_k\left (E(z,t) e^{-i(\omega t-kz)}+E^*(z,t) e^{i(\omega t-kz)}\right).
\label{eq88}
\end{eqnarray}
When substituting these expressions into (\ref{eq83}), the rotating wave approximation \cite{Fleming_2010} is assumed, which consists in neglecting the rapidly oscillating terms containing $e^{\pm 2i\omega t}$. While in (\ref{eq85}) we assume that the amplitudes $R(z,t)$ and $E(z,t)$ are slowly varying functions of their arguments (the slowly varying envelope approximation \cite{Ermolaev_1996}), and neglect $\frac{\partial^2 E}{\partial z^2}$ compared to $k\frac{\partial E}{\partial z}$, as well as $\frac{\partial^2 R}{\partial t^2}$ and $\omega\frac{\partial E}{\partial z}$ compared to $\omega^2 R$. As a result of these approximations (\ref{eq83}) and (\ref{eq85}) are replaced by Maxwell-Bloch equations (note that according to (\ref{eq85}) and (\ref{eq87}) $\boldsymbol{\epsilon}_k\parallel{\bf{d}}$):
\begin{eqnarray} &&
\frac{\partial R}{\partial t}=-i\frac{d}{\hbar}\,EZ,\;\;
\frac{\partial Z}{\partial t}=i\frac{d}{2\hbar}\left (ER^*-E^*R\right),\nonumber \\ &&\frac{\partial E}{\partial z}+\frac{1}{c}\frac{\partial E}{\partial t}=i\frac{2\pi\omega \,n}{c}d\,R,
\label{eq89}
\end{eqnarray}
where $Z=\rho_{ee}-\rho_{gg}$ is the population difference per atom between the excited and ground states. In the case of magnetic dipole transitions, we obtain similar equations with the replacement of $E\to H$ and ${\bf{d}}\to\boldsymbol{\mu}$.

The Maxwell-Bloch equations (\ref{eq89}) describe superradiance as an essentially one-dimensional problem. However, this implicitly assumes one more approximation. If the radius of the cylinder is too small compared to the emitted wavelength $2\pi c/\omega$, diffraction becomes significant and destroys one-dimensionality. On the other hand, at a large cylinder radius, the radiation supports off-axis modes and the transverse variability of the field can no longer be ignored. The trade-off between too much off-axis radiation and too much diffraction is controlled by the so-called Fresnel number:
\begin{equation}
    F=\frac{\omega r^2}{2Lc},
\label{eq90}    
\end{equation}
where $r$ is the radius of the cylinder and $L$ is its length. The validity of (\ref{eq89}) assumes $F\approx 1$ \cite{Gross_1982}.

The Maxwell-Bloch equations (\ref{eq89}) more conveniently can be expressed in terms of dimensionless variables $\tau$, $\xi$ and $\epsilon$ defined in the following way \cite{Ermolaev_1996}
\begin{equation}
 \tau=t\,\Omega,\;\;\xi=z\,\frac{\Omega}{c},\;\;\epsilon=-i\frac{d \,E}{\hbar\Omega},\;\; \Omega^2=\frac{2\pi\omega\,n}{\hbar}\,d^2,
 \label{eq91}
\end{equation}
where the last equation is dictated by the requirement that the last equation in (\ref{eq89}) takes its simplest form. Then (\ref{eq89}) is rewritten as follows
\begin{equation} 
\frac{\partial R}{\partial \tau}=Z\epsilon,\;\;\;
\frac{\partial Z}{\partial \tau}=-\frac{1}{2}\left (\epsilon R^*+\epsilon^*R\right),\;\;\;
\frac{\partial \epsilon}{\partial \tau}+\frac{\partial \epsilon}{\partial \xi}=R\,.
\label{eq92}
\end{equation}
In what follows, we assume that the amplitudes $R$ and $\epsilon$ are real. Furthermore, for a sufficiently small sample, we can neglect the spatial variation of the field $\epsilon$ in (\ref{eq92}). Then (\ref{eq92}) reduces to
\begin{equation} 
\frac{\partial R}{\partial \tau}=Z\epsilon,\;\;\;
\frac{\partial Z}{\partial \tau}=-R\epsilon,\;\;\;
\frac{\partial \epsilon}{\partial \tau}=R\,.
\label{eq93}
\end{equation}
The first two equations show that $R^2+Z^2=B^2$ is a constant, and so we can think of $R$ and $Z$ as the transverse and longitudinal components of the Bloch vector ${\bf{B}}$ with magnitude $B$ and polar angle $\theta(\tau)$:
\begin{equation}
R(\tau)=B\sin{\theta(\tau)},\;\;\; Z(\tau)=B\cos{\theta(\tau)},\;\;\epsilon(\tau)=\frac{d\theta(\tau)}{d\tau},
\label{eq94}
\end{equation}
where the last equation follows, for example, from the first equation (\ref{eq93}). Then the last equation in (\ref{eq93}) takes the form
\begin{equation}
\frac{d^2\theta(\tau)}{d\tau^2}=B\sin{\theta(\tau)},
\label{eq95}
\end{equation}
which coincides with the equation of motion of a rigid pendulum.  Its solution is expressed in terms of elliptic integrals and describes the oscillatory energy flow between atoms and the resonant mode of the field.

To describe superradiance, we must modify the last equation in (\ref{eq93}) to include the emission of radiation from the sample \cite{Ermolaev_1996}. We can obtain the corresponding equation from the energy balance:
\begin{equation}
\frac{1}{2}\hbar\,\omega N\frac{dZ}{dt}+\frac{SL}{8\pi}\,\frac{d|E|^2}{dt}=-\frac{c}{4\pi}\,|E|^2\,S\,,
\label{eq96}
\end{equation}
where $L$ is the length of the sample, $S$ is its cross-section, so that the left-hand side (\ref{eq96}) represents the time derivative of the energy stored in the atoms and the electric field, while the right-hand side represents the Poynting flux from the end faces of the sample. Note that we have to average this flux over the fast oscillations of the electromagnetic field, since the left-hand side concerns quantities in the slow envelope approximation. However, the additional $1/2$ factor introduced by averaging is cancelled out because the energy leakage occurs through both ends of the cylinder.  Using $\frac{d}{dt}=\Omega\frac{d}{d\tau}$, so that $\frac{dZ}{dt}=\Omega\frac{dZ}{d\tau}=-\Omega R\epsilon$, and $|E|=\sqrt{2\pi n\hbar\omega}\,\epsilon$, we easily obtain from (\ref{eq96}) the desired equation \cite{Ermolaev_1996}:
\begin{equation}
\frac{d\epsilon}{d\tau}+\frac{c}{L\Omega}\,\epsilon=R\,.
\label{eq97}
\end{equation}
Since $\epsilon=\frac{d\theta}{d\tau}$, this equation implies
\begin{equation}
\frac{d^2\theta}{dt^2}+\frac{c}{L\Omega}\,\frac{d\theta}{d\tau}=B\sin{\theta},
\label{eq98}
\end{equation}
Which is the equation for a pendulum with friction.

For a small sample size one would expect $\frac{c}{L\Omega}\gg 1$, so the first term in (\ref{eq98}) can be neglected compared to the second (overdamped pendulum). Then such a truncated differential equation (\ref{eq98}) is readily solved. In particular, assuming total initial inversion $B=1$, $\theta(0)=\theta_0\ll 1$, we obtain
\begin{equation}
\frac{L\Omega}{c}\,\tau=\int\limits_{\theta_0}^\theta\frac{d\theta}{\sin{\theta}}=-\int\limits_{\theta_0}^\theta\frac{d\cos{\theta}}{1-\cos^2{\theta}}=
-\left .\frac{1}{2}\ln{\frac{1+cos{\theta}}{1-\cos{\theta}}}\right|_{\theta_0}^\theta.
\label{eq99}
\end{equation}
Solving for $\cos{\theta}$ gives
\begin{equation}
Z=\cos{\theta}=-\tanh{\left [\frac{t-t_0}{2T_{SR}}\right]},\;\;W=\frac{1}{2}NZ=-\frac{N}{2}\tanh{\left[\frac{t-t_0}{2T_{SR}}\right]}\,.
\label{eq100}
\end{equation}
where
\begin{equation}
  T_{SR}=\frac{c}{2L\Omega^2}=\frac{c\hbar}{4\pi\omega \,d^2\,nL}, \;\;t_0=T_{SR}\,\ln{\left(\cot^2{\frac{\theta_0}{2}}\right)}.  
\label{eq101}
\end{equation}
We cannot take the initial tilt angle $\theta_0$ of the Bloch vector to be zero, since a classical rigid pendulum never leaves unstable equilibrium and $t_0\to\infty$ at $\theta_0\to 0$. The second relation in (\ref{eq101}) between the delay time $t_0$ and the tipping angle $\theta_0$ was verified experimentally and was found to be in reasonable agreement with the experimental data \cite{Chen_1999}.

Using (\ref{eq50}), $T_{SR}$ can be expressed in terms of the atomic lifetime constant $\tau_0=1/\gamma$:
\begin{equation}
T_{SR}=\frac{\omega^2\tau_0}{3\pi nLc^2}=\frac{4\pi S}{3N\lambda^2_\omega}\,\tau_0,\;\;\lambda_\omega=\frac{2\pi c}{\omega}.
\label{eq102}
\end{equation}
Comparing (\ref{eq100}) with (\ref{eq16}), we see that
\begin{equation}
\lambda=\frac{3}{4\pi}\,\frac{\lambda_\omega^2}{S}=\frac{3}{2}\,\frac{c}{\omega L}, \label{eq103}
\end{equation}
in reasonable agreement with the first estimate in (\ref{eq78}). The second equality in (\ref{eq103}) follows from the assumption that the Fresnel number $F=\frac{\omega S}{2\pi Lc}=1$.

Let us return to (\ref{eq92}) and again assume that $R$ and $\epsilon$ are real. If we introduce a retarded dimensionless time $\tau_-=\tau-\xi$ and use $(\tau_-,\,\xi)$ as independent variables, so that (note that $\tau=const$ implies $\Delta\tau_-=-\Delta \xi$)
\begin{equation}
\left .\frac{\partial}{\partial\tau}\right |_{\xi=\mathrm{const}}=\left .\frac{\partial}{\partial\tau_-}\right |_{\xi=\mathrm{const}},\;\;\left .\frac{\partial}{\partial\xi}\right |_{\tau=\mathrm{const}}=\left .-\frac{\partial}{\partial\tau_-}\right |_{\xi=\mathrm{const}}+\left .\frac{\partial}{\partial\xi}\right |_{\tau_-=\mathrm{const}},
\label{eq_add_1}
\end{equation}
then (\ref{eq92}) is rewritten in the following way
 \begin{equation} 
\frac{\partial R}{\partial \tau_-}=Z\epsilon,\;\;\;
\frac{\partial Z}{\partial \tau_-}=-R\epsilon,\;\;\;
\frac{\partial \epsilon}{\partial \xi}=R\,.
\label{eq104}
\end{equation}   
By virtue of the first two equations in (\ref{eq104}), the length of the Bloch vector does not change and thus we can again write
\begin{equation}
R(\tau_-,\xi)=B\sin{\theta(\tau_-,\xi)},\;\; Z(\tau_-,\xi)=B\cos{\theta(\tau_-,\xi)}.
\label{eq105}
\end{equation}
Then from the first or second equation in (\ref{eq104}) it follows that
\begin{equation}
\epsilon(\tau_-,\xi)= \frac{\partial \theta(\tau_-,\xi)}{\partial \tau_-},
\label{eq106}
\end{equation}
and the last equation in (\ref{eq104}) is transformed into the sine-Gordon equation:
\begin{equation}
    \frac{\partial^2 \theta(\tau_-,\xi)}{\partial \xi\,\partial\tau_-}=B\sin{\theta(\tau_-,\xi)}\,.
 \label{eq107}   
\end{equation}
The sine-Gordon equation has many applications in physics \cite{Barone_1971}. In the context of superradiance, Burnham and Chiao considered an auto-modeling solution of (\ref{eq107}) that depends on only one variable $q=2\sqrt{\tau_-\xi}$ \cite{Burnham_1969}. Then
\begin{equation}
\frac{\partial}{\partial \tau_-}=\sqrt{\frac{\xi}{\tau_-}}\,\frac{d}{dq},\;\;
\frac{\partial}{\partial \xi}=\sqrt{\frac{\tau_-}{\xi}}\,\frac{d}{dq},
\label{eq_add_2}
\end{equation}
and (\ref{eq104}) reduces to an ordinary differential equation (the Burnham-Chiao equation)
\begin{equation}
 \frac{d^2\theta(q)}{d q^2}+\frac{1}{q}\,\frac{d\theta(q)}{dq}=B\sin{\theta(q)},\;\;q=2\sqrt{\frac{1}{LT_{R}}\,z\left(t-\frac{z}{c}\right)}\,,
 \label{eq108}
\end{equation}
where we introduce $T_R=2T_{SR}$, and $T_{SR}$ is given by (\ref{eq101}). To solve (\ref{eq108}), we need initial values
\begin{equation}
\theta(0)=\theta_0=\frac{2}{\sqrt{N}}\,,\;\;\frac{d\theta}{dq}(0)=0.
\label{eq109}
\end{equation}
Note that we cannot set $q_0=0$, since such a choice corresponds to an unstable equilibrium, and classically the system will remain in this state forever.  The non-zero value of $\theta_0$ simulates quantum fluctuations that initially take the system out of equilibrium. Several approaches to $\theta_0$ can be found in the literature \cite{Vrehen_1979}. We choose the one \cite{Schuurmans_1979,Schuurmans_1982} that has experimental confirmation \cite{Vrehen_1979}.

\section{Hyperfine splitting in hydrogen and positronium}
The spins of the electron and proton in a hydrogen atom can sum either to a total spin of one (triplet state) or to a total spin of zero (singlet state). The energies of the triplet and singlet states differ slightly due to the magnetic interaction between the magnetic moments of the electron and proton. This difference in energy can be estimated as follows \cite{Griffiths_1982}. A proton with a magnetic moment $\boldsymbol{\mu}_p$ creates at the location of an electron with a relative radius vector ${\bf{r}}$ a vector potential ${\bf{A}}$ and, consequently, a magnetic field ${\bf{B}}=\boldsymbol{\nabla}\times{\bf{A}}$, determined by the formulas
\begin{equation}
 {\bf{A}}=\frac{\boldsymbol{\mu}_p\times{\bf{r}}}{r^3},\;\; {\bf{B}}=\boldsymbol{\nabla}\times\left(\frac{\boldsymbol{\mu}_p\times{\bf{r}}}{r^3}\right)=\boldsymbol{\nabla}\times\left(\boldsymbol{\nabla}\times\frac{\boldsymbol{\mu}_p}{r}\right ),
\label{eq110}
\end{equation}
where in the last step we used the equality $$\boldsymbol{\nabla}\frac{1}{r}=-\frac{\bf{r}}{r^3}.$$
Therefore, the interaction Hamiltonian is
\begin{equation}
H_{int}=-\boldsymbol{\mu}_e\cdot\mathbf{B}= -\boldsymbol{\mu}_e\cdot \boldsymbol{\nabla}\times\left(\boldsymbol{\nabla}\times\frac{\boldsymbol{\mu}_p}{r}\right )= -\boldsymbol{\mu}_e\cdot\boldsymbol\nabla\left(\boldsymbol\nabla\cdot\frac{\boldsymbol{\mu}_p}{r}\right )+ \boldsymbol{\mu}_e\cdot\boldsymbol{\mu}_p\,\boldsymbol\nabla^2\frac{1}{r}.
\label{eq111}
\end{equation}
Now we use the identities \cite{Frahm_1983}
\begin{equation}
\boldsymbol\nabla^2\left(\frac{1}{r}\right)=-4\pi\delta({\bf{r}}),\;\;\frac{\partial^2}{\partial r_i\partial r_j}\left(\frac{1}{r}\right)=-\frac{4\pi}{3}\,\delta_{ij}\,\delta({\bf{r}})+\frac{3r_ir_j-r^2\delta_{ij}}{r^5},
\label{eq112}
\end{equation}
and finally obtain
\begin{equation}
H_{int}=\frac{r^2\boldsymbol{\mu}_e\cdot\boldsymbol{\mu}_p-3\left(\boldsymbol{\mu}_e\cdot{\bf{r}}\right)\left(\boldsymbol{\mu}_p\cdot{\bf{r}}\right)}{r^5} -\frac{8\pi}{3}\, \boldsymbol{\mu}_e\,\cdot\boldsymbol{\mu}_p \;\delta({\bf{r}}).
\label{eq113}
\end{equation}
Note that the second equality in (\ref{eq112}) holds as an equality between generalized functions only for test functions that are smooth at the origin \cite{Franklin_2010,Soliverez_1980,Milford_1960}. The interaction Hamiltonian (\ref{eq113}) can be obtained in a less mathematical and more physical way if the magnetic moment of the nucleus is modeled as a rotating uniformly charged sphere whose radius tends to zero \cite{Griffiths_1982,Tinkham_1964}.

Because of this magnetic interaction, in the first order of perturbation theory, the energy of the ground state $1S$ is shifted by the amount
\begin{equation}
\Delta E_{J,M}=\int\Psi^*\hat H_{int}\Psi\,d{\bf{r}},\;\;\Psi=\frac{1}{\sqrt{\pi a_B^3}}e^{-\frac{r}{a_B}}\,|J,M\rangle,\;\;a_B=\frac{\hbar^2}{m_ee^2},
\label{eq114}
\end{equation}
where $a_B$ is the Bohr radius, and $|J,M\rangle$ denotes the spin part of the wave function with $J=1$ or $J=0$. The first term in (\ref{eq113}) gives a zero contribution since the angular integral with this term vanishes. Therefore, since the quantum operators
\begin{equation}
\hat{\boldsymbol{\mu}}_p=g_p\,\frac{e}{2M_pc}\,\hat{{\bf{S}}}_p,\;\; \hat{\boldsymbol{\mu}}_e=-g_e\,\frac{e}{2m_ec}\,\hat{{\bf{S}}}_e,\;\; \hat{{\bf{S}}}_p\cdot\hat{{\bf{S}}}_e=\frac{1}{2}\left(\hat{{\bf{J}}}^2-\hat{{\bf{S}}}_p^2-\hat{{\bf{S}}}_e^2\right ),
\label{eq115}
\end{equation}
where $g_p\approx 5.5857$ and $g_e\approx 2.0023$ are the $g$-factors of the proton and electron in their gyromagnetic ratios, $\hat{{\bf{J}}}=\hat{{\bf{S}}}_p+\hat{{\bf{S}}}_e$ is the total momentum operator of the hydrogen atom in the $1S$ state, and
\begin{equation}
\hat{{\bf{J}}}^2 \,|J,M\rangle=\hbar^2J(J+1)\,|J,M\rangle,\;\;\hat{{\bf{S}}}_p^2\,|J,M\rangle=\hat{{\bf{S}}}_e^2\,|J,M\rangle=\frac{3}{4}\hbar^2,
\label{eq116}
\end{equation}
we get
\begin{equation}
 \Delta E_{J,M}=\frac{g_eg_p}{3a_B^3}\,\frac{e^2\hbar^2}{m_eM_pc^2}\left[J(J+1)-\frac{3}{2}\right]=\frac{4g_eg_p}{3}\,\frac{R^2}{M_pc^2}\left[J(J+1)-\frac{3}{2}\right],
 \label{eq117}
\end{equation}
where $R=\frac{e^2}{2a_B}\approx 13.6$~eV is the binding energy of the hydrogen atom ground state. 

As we can see, $\Delta E_{J,M}$ does not depend on $M$, and the hyperfine splitting $\Delta E_{hyd}=E_{1,M}-E_{0,0}$ in the ground state of the hydrogen atom is equal to \cite{Griffiths_1982}
\begin{equation}
 \Delta E_{hyd}=\frac{8g_eg_p}{3}\,\frac{R^2}{M_pc^2}\approx 5.884\cdot 10^{-6}~\mathrm{eV}.
 \label{eq118}
\end{equation}
The frequency of the photon emitted in the triplet-singlet transition is $\nu_{hyd}=\Delta E_{hyd}/h\approx 1420$~MHz and corresponds to the famous hydrogen line $\lambda_{hyd}=c/\nu\approx 21$~cm. The existence of this line and the possibility of its detection were predicted by Van de Hulst in 1945 \cite{VanDeHulst_1982}. Since its first detection by Ewen and Purcell \cite{Ewen_1951} in 1951, it has become an indispensable tool in radio astronomy and cosmology \cite{Mesinger_2019,Pritchard_2012}.

In the case of positronium, $g_p/M_p$ should be replaced by $g_e/m_e$ in (\ref{eq117}), and the electron mass $m_e$ in $a_B$ should be replaced by the reduced mass $m_e/2$, which will effectively change $a_B$ to $2a_B$. As a result, we obtain
\begin{equation}
\Delta E_{pos}^{(1)}=\frac{g_e^2}{3}\,\frac{R^2}{m_ec^2}.
\label{eq119}
\end{equation}
However, unlike the hydrogen atom, the electron and positron in the triplet state of positronium (ortho-positronium) can annihilate to form a virtual photon, and then this photon is converted into another pair with the same total, but not the same relative momentum. The effective interaction Hamiltonian corresponding to such a quantum electrodynamical process for otho-positronium $1S$ ground state is \cite{Bethe_1957}
\begin{equation}
\hat H_{pair}=2\pi\left(\frac{e\hbar}{m_ec}\right)^2\,\delta({\bf{r}}),
\label{eq120}
\end{equation}
and this creates an additional energy difference between the triplet and singlet states (such an interaction is impossible for the singlet state para-positronium, since the conservation of $C$-parity in electromagnetic interactions prohibits its annihilation into one virtual photon):
\begin{equation}
\Delta E_{pos}^{(2)}=\frac{1}{\pi(8a_B^3)}\,2\pi\left(\frac{e\hbar}{m_ec}\right)^2=\frac{R^2}{m_ec^2}.
\label{eq121}
\end{equation}
Therefore, the total triplet-singlet splitting $\Delta E_{pos}=\Delta E_{pos}^{(1)}+\Delta E_{pos}^{(2)}$ for positronium is equal to
\begin{equation}
\Delta E_{pos}=\left(\frac{g_e^2}{3}+1\right)\,\frac{R^2}{m_ec^2}\approx \frac{7}{3}\,\frac{R^2}{m_ec^2}=\frac{7}{6}\,\alpha^2R\approx 8.45\cdot 10^{-4}~\mathrm{eV}.
\label{eq122}
\end{equation}
The experimental value is \cite{Griffiths_1982} $\Delta E_{pos}\approx 8.411\cdot 10^{-4}~\mathrm{eV}$. The relative discrepancy of about 0.5\% (as well as the discrepancy of about 0.2\% in the case of the hydrogen atom) is due to unaccounted higher-order quantum electrodynamical corrections \cite{Griffiths_1982}. The corresponding triplet-singlet spin-flip transition frequency in positronium is $\nu_{pos}\approx 203.4$~GHz with wavelength $\lambda_{pos}\approx 1.474$~mm. 

Interestingly, Kardashev argued that the window around the positronium spin-flip line is the optimal wavelength region for communication with extraterrestrial intelligence \cite{Kardashev_1979}. A preliminary search for possible extraterrestrial intelligence broadcasting on this frequency yielded negative results \cite{Steffes_1994,Mauersberger_1996}. However, the millimeter and sub-millimeter ranges remain virtually unexplored in SETI surveys \cite{Mason_2024}. Almost the same frequency of 203~GHz is characteristic of the rotational molecular line of $H_2\,^{18}O$, the emission of which from hot molecular clouds has already been detected \cite{Kardashev_1979,Jacq_1988}.

A few words about  how (\ref{eq120}) was obtained \cite{Berestetskii_1982}. The idea is to find the quantum electrodynamic (QED) amplitude corresponding to the annihilation of an electron and positron into a virtual photon by subsequent re-creation of the pair, and interpret it as being caused by an effective three-dimensional potential \cite{Adkins_2022}. The QED amplitude (one of the two amplitudes contributing to Bhabha scattering) is 
\begin{equation}
{\cal M}_{ann}=e^2\left[\bar v(p_+)\gamma^\mu u(p_-)\right]\frac{4\pi}{(p_++p_-)^2}\left[\bar u(p_-^\prime)\gamma_\mu v(p_+^\prime)\right],
\label{eq123}
\end{equation}
where $p_-$ and $p_+$ are the 4-momenta of the electron and positron before annihilation, $p_-^\prime$ and $p_+^\prime$ are 4-momenta after annihilation, $u$ and $v$ denote their corresponding Dirac spinors, and $4\pi$ appears in the photon propagator since we are using Gaussian units. Note that we temporarily assume that $c=\hbar=1$, as is customary in most quantum field theory textbooks.

It is impossible to extract the effective potential of the interaction of an electron and a positron from (\ref{eq123}), since this requires bringing (\ref{eq123}) to a form in which the electron spinors $\bar u(p_-^\prime)$ and $u(p_-)$, as well as the positron spinors $\bar v(p_+)$ and $v(p_+^\prime)$ are contracted.  This is achieved by using the Fierz identity \cite{Pal_2015}
\begin{eqnarray} &
 \left[\bar v(p_+)\gamma^\mu u(p_-)\right]\left[\bar u(p_-^\prime)\gamma_\mu v(p_+^\prime)\right]=&
 \nonumber \\ &
 \frac{1}{4}\Big\{  4\left[\bar u(p_-^\prime)u(p_-)\right]\left[\bar v(p_+)v(p_+^\prime)\right]-
 2\left[\bar u(p_-^\prime)\gamma^\mu u(p_-)\right]\left[\bar v(p_+)\gamma_\mu v(p_+^\prime)\right]-&
\nonumber \\ &
 2\left[\bar u(p_-^\prime)\gamma^\mu\gamma_5 u(p_-)\right]\left[\bar v(p_+)\gamma_\mu\gamma_5 v(p_+^\prime)\right]- 
 4\left[\bar u(p_-^\prime)\gamma_5 u(p_-)\right]\left[\bar v(p_+)\gamma_5 v(p_+^\prime)\right]\Big \}
  .\qquad\;&
\label{eq124}
\end{eqnarray}

The electron and positron in positronium are non-relativistic, so $(p_++p_-)^2\approx 4m_e^2$ and since the photon propagator in this approximation  already contains a factor of $1/c^2$ (when $c$ is restored on dimensional grounds), it is sufficient to use the zero-order approximation for Dirac spinors:
\begin{equation}
 u(p_-)\approx \left(\begin{array}{c} \xi \\ 0 \end{array} \right ),\;\;\;
 v(p_+)\approx \left(\begin{array}{c} 0 \\ \zeta  \end{array} \right ),
\label{eq125}
\end{equation}
where $\xi$ and $\zeta$ are three-dimensional spinors for the electron and positron, respectively. Using the Dirac representation for the $\gamma$-matrices and (\ref{eq124}), we can show after simple algebraic calculations that (\ref{eq123}) takes the form
\begin{equation}
 {\cal M}_{ann}=-\frac{e^2\pi}{2m_e^2}\Big[3\,\xi^{\prime\, +}\xi\,\zeta^+\zeta^\prime+\xi^{\prime\, +}\boldsymbol{\sigma}\,\xi\cdot\zeta^+\boldsymbol{\sigma}\,\zeta^\prime\Big]=-\xi^{\prime\,+}_\alpha\zeta^+_\beta \,V_{\alpha\beta,\gamma\delta}\,\xi_\gamma \zeta^\prime_\delta,
 \label{eq126}
\end{equation}
where the effective interaction operator $V_{\alpha\beta,\gamma\delta}$ was introduced from the requirement that it yield the same matrix element as the Hermitian Gupta ${\cal K}$-matrix related to the unitary $S$ matrix by \cite{Gupta_1964}
\begin{equation}
S=\frac{1-\frac{i}{2}{\cal K}}{1+\frac{i}{2}{\cal K}},
\label{eq127}
\end{equation}
in particular, at the first non-vanishing order ${\cal K}=-{\cal M}$.

We can express the momentum-space effective interaction operator in the matrix form:
\begin{equation}
V_{pair}=\frac{e^2\pi}{2m_e^2}\,\left (3+\boldsymbol{\sigma}^{(-)}\cdot \boldsymbol{\sigma}^{(+)}\right),
\label{eq128}
\end{equation}
Assuming that $\boldsymbol{\sigma}^{(-)}$ acts only on electron spinors, and $\boldsymbol{\sigma}^{(+)}$ acts only on positron spinors. The effective potential in the configuration space is obtained by the Fourier transform
\begin{equation}
 V_{pair}({\bf{r}})=\frac{1}{(2\pi\hbar)^3}\int e^{\frac{i}{\hbar}{\bf{q}}\cdot{\bf{r}}}\,V_{pair}({\bf{q}}),
 \label{eq129}
\end{equation}
and since $V_{pair}({\bf{q}})$ is actually independent of the momentum ${\bf{q}}$, integration will yield a $\delta$-function, and we finally obtain (we have restored $\hbar$ and $c$ on demensional grounds)
\begin{equation}
V_{pair}({\bf{r}})=\frac{\pi}{2}\left(\frac{e\hbar}{m_ec}\right)^2\left (3+\boldsymbol{\sigma}^{(-)}\cdot \boldsymbol{\sigma}^{(+)}\right)\,\delta({\bf{r}}).
 \label{eq130}
\end{equation}
Since $\boldsymbol{\sigma}^{(-)}=\frac{2}{\hbar}\,\hat{\bf{S}}^{(-)}$, $\boldsymbol{\sigma}^{(+)}=\frac{2}{\hbar}\,\hat{\bf{S}}^{(+)}$, we easily find
$$\langle J,M|\left (3+\boldsymbol{\sigma}^{(-)}\cdot \boldsymbol{\sigma}^{(+)}\right)|J,M\rangle=2J(J+1),$$
where $J$ is the total momentum of positronium in the $1S$ state (coinciding with the total spin), and we see that (\ref{eq130}) is equivalent to (\ref{eq120}).

Let us estimate the lifetime of a spontaneous hyperfine transition from a triplet to a singlet state \cite{Harris_1972,Kuzmak_2024}. The corresponding transition rate is determined by Fermi's golden rule \cite{Stedman_1971}
\begin{equation}
\gamma=\frac{2\pi}{\hbar}\sum\limits_{{\bf{k}},\sigma}\left|\langle f|\hat{H}_{int}|i\rangle\right|^2\delta(E_f-E_i),
\label{eq131}
\end{equation}
where the interaction Hamiltonian is the magnetic dipole part of (\ref{eq49}) with
\begin{equation}
\boldsymbol{\mu}=\frac{eg_p}{2M_pc}\,\hat{{\bf{S}}}_p\,-\, \frac{eg_e}{2m_ec}\,\hat{{\bf{S}}}_e=
-\frac{eg_e}{2m_ec}\left( \hat{{\bf{S}}}_e-\frac{m_eg_p}{M_pg_e}\,\hat{{\bf{S}}}_p\right ).
\label{eq132}
\end{equation}
Of course, for hydrogen, due to the large difference in mass between the electron and the proton, the proton's contribution to (\ref{eq132}) can be neglected. However, keeping in mind the further application to positronium, we will retain it.

For such an interaction Hamiltonian, we have already calculated the decay rate and obtained as a result (\ref{eq55}). So, we just need to calculate $|\langle g|\boldsymbol{\mu}|e\rangle|^2$ for $\boldsymbol{\mu}$ from (\ref{eq132}). Using
\begin{equation}
\left |\langle \uparrow |\boldsymbol{\sigma}|\downarrow \rangle \right |^2=\left |\langle \downarrow |\boldsymbol{\sigma}|\uparrow \rangle \right |^2=2,\;\; \left |\langle \uparrow |\boldsymbol{\sigma}|\uparrow \rangle \right |^2=1,\;\;\langle \uparrow |\boldsymbol{\sigma}|\uparrow \rangle =-\langle \downarrow |\boldsymbol{\sigma}|\downarrow \rangle,
\label{eq133}
\end{equation}
we find that the decay rate does not depend on the magnetic quantum number $M$ of the initial triplet state:
\begin{equation}
   |e\rangle=|1,M\rangle=\left \{ \begin{array}{l} |\uparrow \Uparrow\rangle,\;\;\mathrm{if}\;\;M=1, \\
    \frac{1}{\sqrt{2}}\left (|\uparrow\Downarrow\rangle+|\downarrow\Uparrow\rangle\right),\;\;\mathrm{if}\;\;M=0,\\ |\downarrow \Downarrow\rangle,\;\;\mathrm{if}\;\;M=-1,
    \end{array}\right .\ \;|g\rangle=|0,0\rangle=\frac{1}{\sqrt{2}}\left (|\uparrow\Downarrow\rangle-|\downarrow\Uparrow\rangle\right),
\end{equation}
and equals to \cite{Kuzmak_2024}
\begin{equation}
\gamma_{hyp}=\frac{\hbar e^2g_e^2}{12m_e^2c^5}\left (1+\frac{m_eg_p}{M_pg_e}\right)^2 \omega^3=\frac{\alpha g_e^2}{12}\left (1+\frac{m_eg_p}{M_pg_e}\right)^2\left(\frac{\hbar \omega}{m_ec^2}\right)^2\omega,
\label{eq135}
\end{equation}
with $\omega=\frac{\Delta E_{hyp}}{\hbar}$. Therefore, for hydrogen
\begin{equation}
 \gamma_{hyd}\approx\frac{1}{3}\alpha \left(\frac{\hbar \omega_{hyd}}{m_e c^2}\right)^2\omega_{hyd}\approx 2.88\cdot 10^{-15}~\mathrm{s}^{-1},\;\;\;\tau_{hyd}=\frac{1}{\gamma_{hyd}}\approx 3.5\cdot10^{14}~\mathrm{s},
 \label{eq136}
\end{equation}
while for positronium, when $\frac{g_p}{M_p}\to \frac{g_e}{m_e}$ in (\ref{eq135}), and when the experimental value $\Delta E_{pos}\approx 8.411\cdot 10^{-4}~\mathrm{eV}$ is used, 
\begin{equation}
 \gamma_{pos}\approx\frac{4}{3}\alpha \left(\frac{\hbar \omega_{pos}}{m_e c^2}\right)^2\omega_{pos}\approx 3.37\cdot 10^{-8}~\mathrm{s}^{-1},\;\;\;\tau_{pos}=\frac{1}{\gamma_{pos}}\approx 3\cdot 10^{7}~\mathrm{s},
 \label{eq137}
\end{equation}
It is noteworthy that the theoretical value $\gamma_{pos}\approx 3.37\cdot 10^{-8}~\mathrm{s}^{-1}$ is in good agreement with the experimental value $\gamma=3.1^{+1.6}_{-1.2}\,\cdot10^{-8}~\mathrm{s}^{-1}$ obtained in the first direct measurement of the hyperfine transition of the ground state positronium \cite{Yamazaki_2012}. It is perhaps worth mentioning that the theoretical value given in \cite{Yamazaki_2012} is taken from \cite{Wallyn_1996}, which in turn is based on formula (2-132) in \cite{Lang_1980}. Although the numerical value given in \cite{Wallyn_1996} is correct and agrees with our result, the analytical formulas in both \cite{Wallyn_1996} and \cite{Lang_1980} lack the factor 4/3, without which the correct factor $10^{-42}$ in (2-132) of \cite{Lang_1980} cannot be obtained.

\begin{acknowledgments}
We thank the anonymous reviewer for constructive comments that helped improve the readability of the manuscript.
\end{acknowledgments}
\bibliography{DsupR.bib}
\end{document}